\documentclass[fleqn,usenatbib]{mnras}

\usepackage[T1]{fontenc}

\DeclareRobustCommand{\VAN}[3]{#2}
\let\VANthebibliography\thebibliography
\def\thebibliography{\DeclareRobustCommand{\VAN}[3]{##3}\VANthebibliography}

\usepackage{amsmath,amstext,physics}
\usepackage{float}
\usepackage[export]{adjustbox}
\usepackage{graphicx}
\usepackage{printlen,enumitem}
\usepackage{txfonts} 
\usepackage[figure,figure*]{hypcap} 
\usepackage{sidecap}
\usepackage{comment}

\newcommand{\nf}{{\bf \textsc{no-wind}}}
\newcommand{\mf}{{\bf \textsc{AGN-wind}}}
\newcommand\msun[1]{{\rm M}_{\odot}}
\newcommand{\Msun}{{\rm M}_{\odot}}
\newcommand{\Msunh}{{\rm M}_{\odot}\,h^{-1}}
\newcommand{\Msunyr}{{\rm M}_{\odot}\,{\rm yr}^{-1}}
\newcommand{\kms}{{\rm km}\,{\rm s}^{-1}}

\newcommand{\eff}{$\epsilon_{\rm ff}$}
\newcommand{\sigmasfr}{$\Sigma_{\rm SFR}$}

\title[Local positive AGN feedback in total negative]{Local positive feedback in the overall negative: the impact of quasar winds on star formation in the FIRE cosmological simulations}

\author[J. Mercedes-Feliz et al.]{Jonathan~Mercedes-Feliz,$^{1}$\thanks{E-mail: jonathan.mercedes\_feliz@uconn.edu}
Daniel Angl{\'e}s-Alc{\'a}zar,$^{1,2}$
Christopher C. Hayward,$^{2}$
Rachel K. Cochrane,$^{2,3}$\newauthor
Bryan A. Terrazas,$^{4}$
Sarah Wellons,$^{5,8}$
Alexander J. Richings,$^{6,7}$
Claude-Andr{\'e} Faucher-Gigu{\`e}re,$^{8}$\newauthor
Jorge Moreno,$^{9}$
Kung Yi Su,$^{2,10,14}$
Philip F. Hopkins,$^{11}$
Eliot Quataert,$^{12}$
and Du{\v s}an Kere{\v s}$^{13}$
\\
$^{1}$Department of Physics, University of Connecticut, 196 Auditorium Road, U-3046, Storrs, CT 06269-3046, USA\\
$^{2}$Center for Computational Astrophysics, Flatiron Institute, 162 5th Avenue, New York NY 10010, USA\\
$^{3}$Harvard-Smithsonian Center for Astrophysics, 60 Garden St, Cambridge, MA 02138, USA\\
$^{4}$Columbia Astrophysics Laboratory, Columbia University, 550 West 120th Street, New York, NY 10027, USA\\
$^{5}$Department of Astronomy, Van Vleck Observatory, Wesleyan University, 96 Foss Hill Drive, Middletown, CT 06459, USA\\
$^{6}$E. A. Milne Centre for Astrophysics, Department of Physics and Mathematics, University of Hull, Cottingham Road, Hull, HU6 7RX, UK\\
$^{7}$DAIM, University of Hull, Cottingham Road, Hull, HU6 7RX, UK\\
$^{8}$CIERA and Department of Physics and Astronomy, Northwestern University, 1800 Sherman Ave., Evanston, IL 60201, USA\\
$^{9}$Department of Physics and Astronomy, Pomona College, 333 N. College Way, Claremont, CA 91711, USA\\
$^{10}$Department of Astronomy, Columbia University, 550 West 120th Street, New York, NY 10027, USA\\
$^{11}$TAPIR, Mailcode 350-17, California Institute of Technology, Pasadena, CA 91125, USA\\
$^{12}$Department of Astrophysical Sciences, Princeton University, Princeton, NJ 08544, USA\\
$^{13}$Department of Physics, Center for Astrophysics and Space Sciences,University of California San Diego,
9500 Gilman Drive, La Jolla, CA 92093, USA\\
$^{14}$Black Hole Initiative, Harvard University, 20 Garden St., Cambridge, MA 02138, USA
}

\date{Accepted XXX. Received YYY; in original form ZZZ}

\pubyear{2023}

\begin{document}
\label{firstpage}
\pagerange{\pageref{firstpage}--\pageref{lastpage}}
\maketitle

\begin{abstract}
Negative feedback from accreting supermassive black holes is considered crucial in suppressing star formation and quenching massive galaxies. However, several models and observations suggest that black hole feedback may have a positive effect, triggering star formation by compressing interstellar medium gas to higher densities. We investigate the dual role of black hole feedback using cosmological hydrodynamic simulations from the Feedback In Realistic Environments (FIRE) project, incorporating a novel implementation of hyper-refined accretion-disc winds. Focusing on a massive, star-forming galaxy at $z \sim 2$ ($M_{\rm halo} \sim 10^{12.5}\,{\rm M}_{\odot}$), we demonstrate that strong quasar winds with a kinetic power of $\sim10^{46}$ erg/s, persisting for over 20\,Myr, drive the formation of a central gas cavity and significantly reduce the surface density of star formation across the galaxy's disc. The suppression of star formation primarily occurs by limiting the availability of gas for star formation rather than by evacuating the pre-existing star-forming gas reservoir (preventive feedback dominates over ejective feedback). Despite the overall negative impact of quasar winds, we identify several potential indicators of local positive feedback, including: (1) the spatial anti-correlation between wind-dominated regions and star-forming clumps, (2) higher local star formation efficiency in compressed gas at the edge of the cavity, and (3) increased contribution of outflowing material to local star formation. Moreover, stars formed under the influence of quasar winds tend to be located at larger radial distances. Our findings suggest that both positive and negative AGN feedback can coexist within galaxies, although the local positive triggering of star formation has a minor influence on global galaxy growth.
\end{abstract}




\section{Introduction}
A copious amount of evidence show that most galaxies host a massive black hole (BH) at their centre, with the mass of the BH strongly correlating with global galaxy properties \citep{Magorrian1998,Bennert2011,Kormendy&Ho2013,McConnell&Ma2013,Reines2015,Graham2016,Shankar2020}.
Actively accreting BHs in active galactic nuclei (AGN) can impact the host galaxy through a variety of feedback mechanisms, including fast accretion-driven winds \citep{Faucher-Giguere2012,Zubovas2012,Tombesi2013,Nardini2015}, galaxy-scale outflows \citep{feruglio2010,sturm2011,Greene2012,cicone2014,Garcia-Burillo2014,Zakamska2014,Circosta2018,Wylezalek2020,RamosAlmeida2022}, and large scale jets \citep{Fabian2012}. 
Observational constraints on the efficiency of AGN feedback suggest that massive BHs may play a key role in galaxy evolution by injecting energy and momentum into the interstellar medium (ISM) and circumgalactic medium (CGM) of galaxies \citep{Hopkins2010,AlexanderHickox2012,Fabian2012,Alatalo2015,Wylezalek2016,Fiore2017,Harrison2017,harrison2018}. Most galaxy formation models indeed require some form of {\it negative} AGN feedback to eject existing ISM gas from massive galaxies and/or prevent CGM gas from cooling and accreting onto the galaxy, suppressing star formation and quenching massive galaxies, regulating their sizes and central densities, and reproducing the observed bimodality in galaxy colours \citep{DiMatteo2005,Baldry2006,Bower2006,Croton2006,Dubois2012,Silk2012,Somerville&Dave2015,Choi2018,Dave2019,Wellons2023}.

However, recent observations suggest that AGN feedback could also have {\it positive} effects, by triggering star formation, as opposed to suppressing it, in galaxies. 
Direct observational evidence of AGN outflows triggering star formation is slim, but examples exist where star formation seems to occur within the outflow itself. 
\citet{maiolino2017} studied a merging system that hosts an obscured AGN with a prominent outflow and found that multiple optical and near-infrared (IR) diagnostics of the outflowing gas are consistent with star formation within the outflow, where the inferred star formation rate (SFR) can exceed $15\,\Msunyr$~and account for $\sim$25\% of the total SFR in the system.
Analyzing over 2,500 galaxies in MaNGA DR2, \citet{gallagher2019} identified a subsample of 37 galaxies with outflows, of which $\sim$30\%~show signs of star formation within the outflowing gas, ranging from 0.1--$1\,\Msunyr$~and contributing 5--30\%~of the total SFR in the galaxy.
Positive and negative AGN feedback do not necessarily act against one another, and some observations suggest that their effects could simultaneously be present within the same galaxy. \citet{cresci2015b} used SINFONI near-IR integral field spectroscopy of an obscured quasar at $z \sim 1.6$ to show that a prominent outflow traced by [OIII] lines coincides with the location of an empty central cavity surrounded by star-forming regions, suggesting that the outflow is removing gas from the cavity (negative feedback) while triggering star formation at the edge of the cavity (positive feedback).
Additional plausible implications of positive AGN feedback include the alignment of non-thermal radio emission and rest-frame UV continuum emission in radio galaxies, suggesting jet-induced star formation in the host galaxy \citep{Bicknell2000,Zirm2005,Drouart2016}, higher SFR in green-valley galaxies with X-ray detected AGN and far infrared emission \citep{Mahoro2017}, and large scale expanding bubbles powered by jets potentially triggering star formation in other galaxies \citep{Gilli2019}.

Several theoretical models have explored the conditions under which AGN feedback can trigger star formation by compressing interstellar medium gas to higher densities, using analytic calculations \citep{BegelmanCioffi1989,Rees1989,Natarajan1998,King2005,Silk2005,ishibashi2012,silk2013,zubovas2013,nayakshin2014} and hydrodynamic simulations of idealized systems \citep{Gaibler2012,Zubovas2012,bieri2015,bieri2016,zubovas2017}.  
Some of these models propose that AGN feedback triggering of star formation plays a key role driving simultaneous AGN and star formation in galaxies \citep{King2005},
the observed correlation between star formation and AGN luminosities \citep{zubovas2013},
the similarity in the comoving BH accretion rate density and the cosmic star formation history \citep{silk2013}, 
the BH--galaxy scaling relations \citep{nayakshin2014},
the extreme SFRs of high redshift starbursts \citep{Silk2005,Gaibler2012,bieri2015,bieri2016},
the size and structural evolution of massive galaxies \citep{ishibashi2012,Ishibashi2014}, and
the formation of dark matter-deficient dwarf galaxies from swept up gas in the intergalactic medium by quasar outflows \citep[][but see \citealt{Moreno2022}]{Natarajan1998}.
Given the plausible strong implications for galaxy evolution predicted by these idealized models, it is crucial to investigate the impact of positive AGN feedback in more realistic simulations of galaxy formation in a full cosmological context.

Large-volume cosmological hydrodynamic simulations, however, generally do not predict positive AGN feedback scenarios, in part by construction because their subgrid AGN feedback models are implemented to help suppress star formation and regulate the growth of massive galaxies when stellar feedback is not sufficiently strong \citep[see the review by][and references therein]{Somerville&Dave2015}. In this context, the connection between AGN and star formation activity in galaxies is more naturally explained by a common gas supply for star formation and BH growth \citep{Angles-Alcazar2015,Volonteri2015,Angles-Alcazar2017a,Ricarte2019,Thomas2019}, and the extreme SFRs of high redshift galaxies are primarily driven by the systematically higher cosmological gas accretion rate onto halos and/or higher incidence of galaxy mergers \citep{Dave2010,Hayward2013,Narayanan2015}, without requiring AGN feedback-driven triggering of star formation.  
However, cosmological simulations generally lack the resolution to model in detail the interaction of AGN-driven winds or jets with the multi-phase interstellar medium (ISM).

In this paper, we investigate the plausible dual role of AGN feedback in galaxies using high-resolution cosmological zoom-in simulations from the Feedback
In Realistic Environments (FIRE\footnote{\url{http://fire.northwestern.edu}}) project \citep{Hopkins2014,Hopkins2018}, including local stellar feedback by supernovae, stellar winds, and radiation in a multi-phase ISM, and a novel implementation of hyper-refined AGN-driven winds that simultaneously captures their propagation and impact from the inner nuclear region ($\lesssim$10\,pc) to circumgalactic medium (CGM) scales (Angl{\'e}s-Alc{\'a}zar et al., in prep.).
Focusing on a massive, star-forming galaxy near the peak of cosmic activity \citep[$z\sim2$;][]{Madau&Dickinson2014}, we investigate its subsequent evolution over a $\sim$35\,Myr period in simulations with different AGN feedback strengths compared to that of an identical control simulation without AGN feedback. This provides an ideal framework to evaluate any positive versus negative feedback effects on the host galaxy.

The outline of this paper is as follows: \S\ref{sec:methods} provides a brief summary of our methodology and \S\ref{Overview of Simulations} presents an overview of our simulations.  In \S\ref{Negative Impact of AGN feedback} we investigate the global negative impact of AGN feedback on our simulated galaxy while \S\ref{Signatures of (Local) Positive AGN feedback} investigates plausible signatures of positive AGN feedback. In \S\ref{Dependence on AGN wind efficiency} we analyse the impact of AGN winds in simulations that vary the kinetic feedback efficiency. We discuss our results in \S\ref{Discussion} and present our summary and conclusions in \S\ref{Summary and Conclusions}.

\begin{figure*}
\includegraphics[width = 1\textwidth]{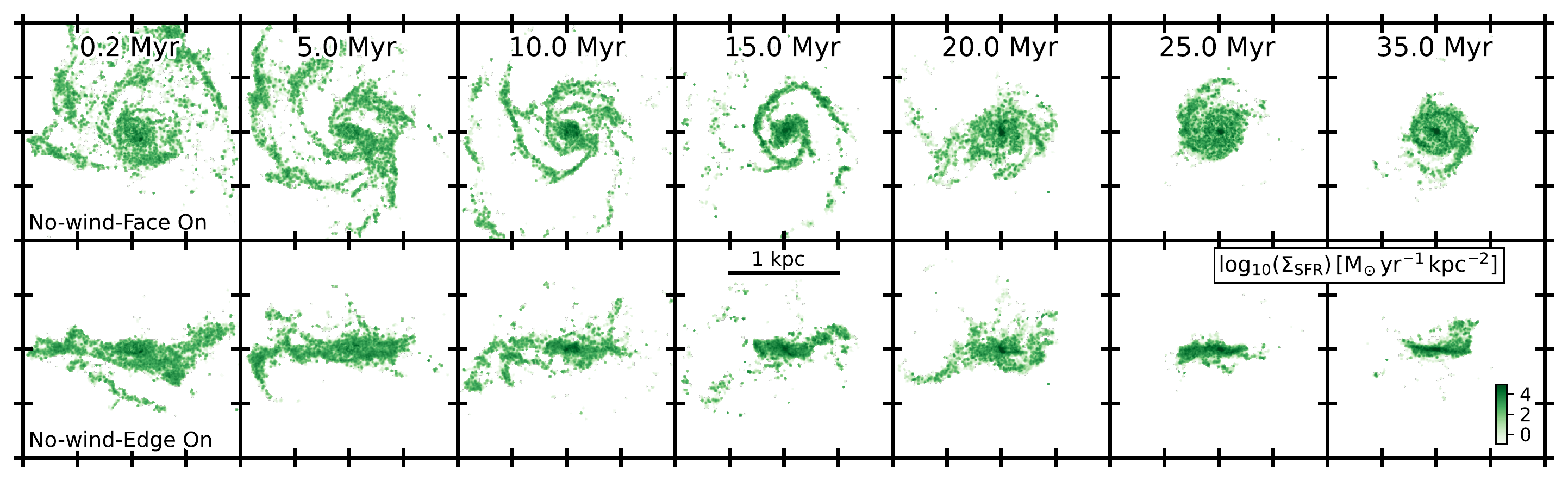}
\includegraphics[width = 1\textwidth]{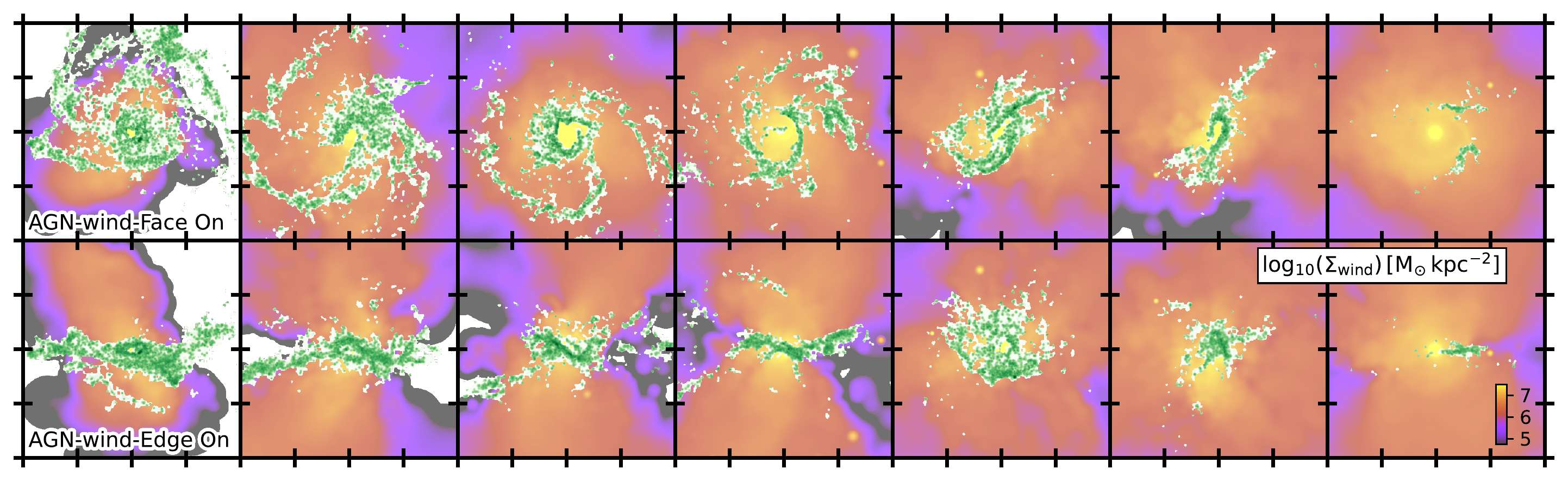}
\vspace*{-5mm}
\caption{\label{fig:SFRall} Projected star formation rate surface density (\sigmasfr) for the central 1\,kpc region of a massive, star-forming galaxy ($M_{\rm star}\sim 10^{11}\,\Msun$, ${\rm SFR}\sim 300\,\Msunyr$) at $z\sim2.28$ for the \nf~(top rows) and \mf~(bottom rows) simulations. In both cases we show edge-on and face-on views along with time evolution (from left to right) for $\sim$35\,Myr since the start ($\Delta t=0$) of the quasar feedback phase in the \mf~simulation. The projected mass density distribution of AGN winds is overlaid in the bottom rows, as indicated by the colour scale. In the absence of AGN-driven winds, the star-forming disc becomes denser and more compact as time progresses. In contrast, AGN winds evacuate star-forming gas from the central region, with the formation of a growing central cavity and global suppression of star formation by the end of the simulation.} 
\end{figure*}

\begin{figure*}
\includegraphics[width = 1\textwidth]{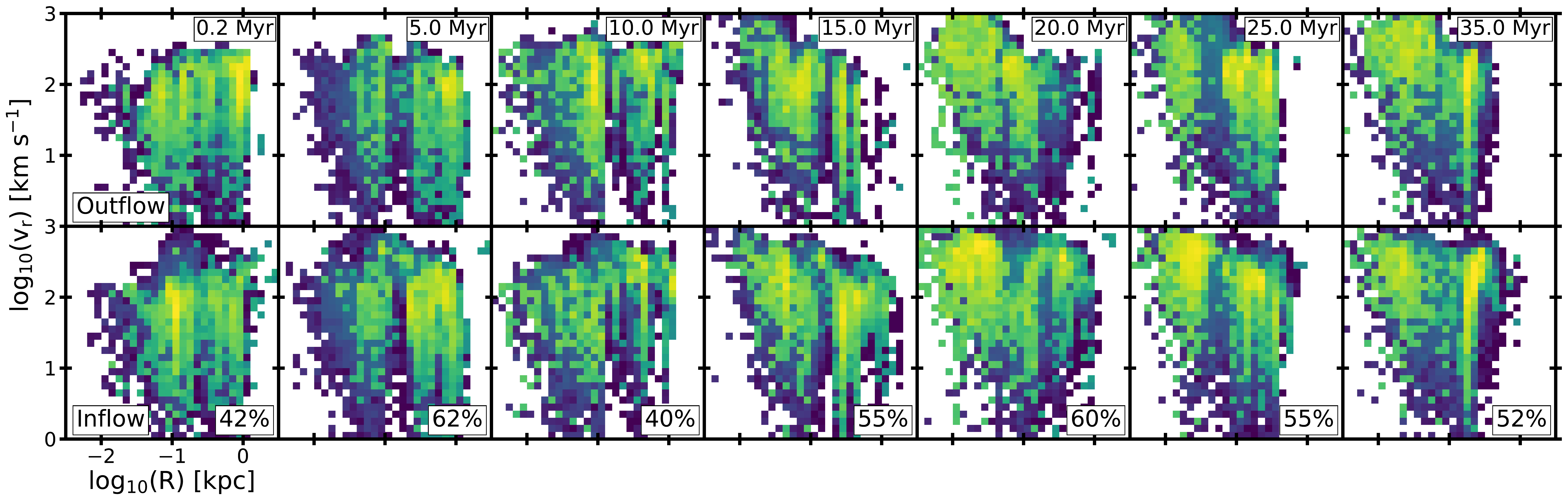}
\includegraphics[width = 1\textwidth]{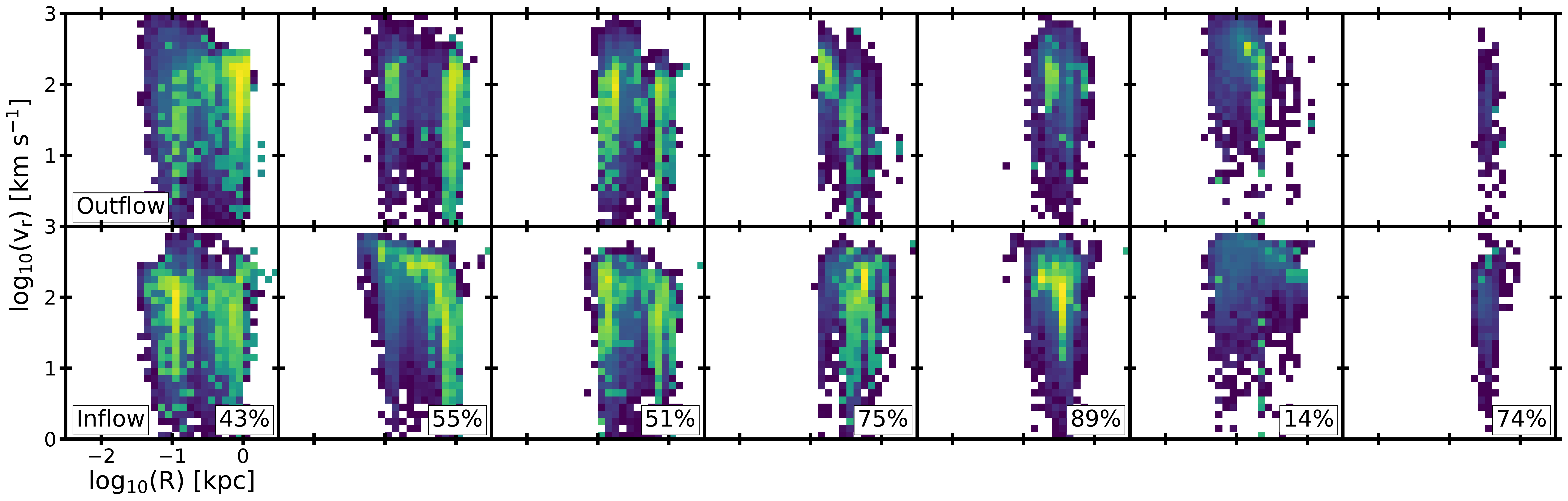}
\vspace*{-5mm}
\caption{\label{fig:RVvsRDall} Two-dimensional, SFR-weighted distribution of radial velocity and radial distance for gas in the central $\sim$3\,kpc for the \nf~(top rows) and \mf~(bottom rows) simulations. Time evolution is shown from left to right for $\sim$35\,Myr (the colour scale is logarithmic and uniform throughout). In both cases we show separately the outflowing (1st and 3rd row) and inflowing (2nd and 4th row) star formation components based on radial velocity. The fraction of total galaxy SFR in the inflowing gas component is indicated in each case. The inflowing and outflowing components for the \nf~case are very similar in their contribution to the SFR, varying anywhere between $\sim$ 40--60\%~of the total SFR in any one of the components. AGN winds introduce stark differences in the amount and distribution of star-forming gas, with the inflowing and outflowing components contributing as much as $\sim$80--90\%~of the total SFR at different times.}
\end{figure*}


\begin{table}
\centering
\begin{tabular}{c| c c c c c} 
    \hline \hline
    Name & $\eta_{k}$ & $\epsilon_{k}$ & $\dot{M}_{\rm w}\,[\Msunyr]$ & $\dot{E}_{\rm w}\,[{\rm erg}\, {\rm s}^{-1}]$ &  Notes\\ [0.05ex] 
    \hline
    \textbf{noAGN} & - & - & - & - &\nf \\ 
    \hline
    \textbf{m0.1e0.5} & 0.1 & 0.005  & 2.22 & $6.29 \times 10^{44}$ & \\
    \hline
    \textbf{m1e5} & 1 & 0.05 & 22.2 & $6.29 \times 10^{45}$ & \\
    \hline
    \textbf{m2e10} & 2 & 0.1 & 44.4 & $1.26 \times 10^{46}$ & \mf \\
    \hline
    \textbf{m4e20} & 4 & 0.2 & 88.8 & $2.52 \times 10^{46}$ & \\
    \hline
    \textbf{m10e50} & 10 & 0.5  & 222 & $6.29 \times 10^{46}$ & \\ [1ex] 
    \hline
    \end{tabular}
\caption{{Simulation parameters: (1) Name: simulation designation. (2) $\eta_{k} \equiv \dot{M}_{w}/\dot{M}_{\rm BH}$: mass loading factor. (3) $\epsilon_{k} \equiv \dot{E}_{\rm w}/L_{\rm bol}$: kinetic feedback efficiency. (4) $\dot{M}_{w}$: mass outflow rate in winds. (5) $\dot{E}_{w}$: kinetic energy injection rate. (6) Notes.}}
\label{table:models}
\end{table}

\section{Methods}
\label{sec:methods}
Angl{\'e}s-Alc{\'a}zar et al. (in prep.) fully describes the simulations and methodology that we implement, which we briefly summarize below.

\subsection{FIRE-2 galaxy formation model} \label{subsec:FIRE2model}

The simulations are part of the FIRE-2 project, an updated implementation of the original FIRE simulations. 
The GIZMO\footnote{\url{http://www.tapir.caltech.edu/~phopkins/Site/GIZMO.html}} solver is used in its meshless finite mass (MFM) mode, which implements a Lagrangian Godunov formulation with many of the benefits of particle and grid-based methods \citep{Hopkins2015gizmo}.
We include cooling and heating from $T=10-10^{10}\,{\rm K}$; star formation in locally self-gravitating, dense ($n_{\rm H}>1000\,{\rm cm}^{-3}$), molecular, and Jeans-unstable gas; and stellar feedback from OB \&\ AGB mass-loss, Type Ia \&\ II Supernovae (SNe), and multi-wavelength photo-heating and radiation pressure \citep{Hopkins2018}. Each star particle represents a population of stars with known mass, age, and metallicity, with all stellar feedback quantities and their time dependence taken from the \textsc{starburst99} population synthesis model \citep{Leitherer1999}.

A thorough description of the star formation model in FIRE-2 is outlined in \S~2.4 of \citet{Hopkins2018} but we briefly summarize below the conditions that gas elements are required to satisfy in order to be eligible for star formation:
\begin{itemize}
     \item \textbf{Self-gravitating:} We require the gravitational potential energy be larger than the thermal plus kinetic energy within the resolution scale, using the sink-particle criterion of \citet{hopkins2013d}. This allows for identifying gas which is collapsing under self-gravity at the resolution scale while not allowing unbound material to form stars, such as tidally unbound gas or dense gas in strong shocks.
     
     \item \textbf{Self-shielding:} The self-shielded, molecular fraction of gas elements is estimated based on the \citet{Krumholz2011} approach and determines the amount of gas eligible for star formation, which depends on the local column density using the Sobolev approximation, gas metallicity, and the ability of gas to cool to low temperatures.
     
     \item \textbf{Jeans Unstable:} To ensure the coherent collapse of resolved, massive self-gravitating objects, we impose a requirement that the thermal Jeans mass remains below either the particle mass or $10^{3}\,M_{\odot}$ in each element.
     
     \item \textbf{Dense:} In order to avoid incorrect application of the aforementioned criteria, an additional check is performed to ensure that the gas density ($n_{\rm H}$) exceeds a critical value ($n_{\rm crit} = 1000\,{\rm cm}^{-3}$). This requirement restricts star formation to dense molecular clouds that emerge from the background disk through fragmentation.
\end{itemize} 
Gas elements that satisfy all these requirements form stars at a rate $\dot{\rho}_\star = \rho_{\rm mol}/t_{\rm ff}$, where $\rho_{\rm mol}$ is the molecular gas density and $t_{\rm ff}$ is the local free-fall time.

\subsection{Initial conditions} \label{subsec:initialconditions}
Our initial conditions are derived from snapshots of pre-existing FIRE-2 simulations, and adopted to include AGN-driven winds in our new simulations. We focus on the massive FIRE-2 halo \textbf{A4} from \citet{Angles-Alcazar2017c}, with $M_{\rm halo}\sim 10^{12.5}\,{\rm M}_{\odot}$ at $z=2$ and evolved down to $z=1$ including on-the-fly BH growth driven by gravitational torques \citep{Hopkins&Quataert2011,Angles-Alcazar2013,Angles-Alcazar2015,Angles-Alcazar2017a} but not including BH feedback. 
The new simulations with AGN winds adopt the same baryonic mass resolution $m_{\rm b}=3.3\times 10^{4}\,{\rm M}_{\odot}$ and force softenings $\epsilon_{\rm gas}^{\rm min}=0.7\,{\rm pc}$, $\epsilon_{\star}=7\,{\rm pc}$ and $\epsilon_{\rm DM}=57\,{\rm pc}$ for the gas (minimum adaptive force softening), stellar, and dark matter components. We assume a $\Lambda$CDM cosmology with parameters $H_{0}=69.7\,{\rm km}\,{\rm s}^{-1}\,{\rm Mpc}^{-1}$, $\Omega_{\rm M}=1-\Omega_{\Lambda}=0.2821$, $\Omega_{\rm b}=0.0461$, $\sigma_{8}=0.817$, and $n_{\rm s}=0.9646$ \citep{Hinshaw2013}.

We choose to inject AGN winds in the new simulations at the $z= 2.28$ snapshot, which will be referenced hereafter as $\Delta t=0\,{\rm Myr}$. 
At this time, the galaxy is undergoing a strong starburst phase which will lead to the formation of an overcompact and overdense stellar system because stellar feedback is no longer able to regulate star formation  \citep[][Angl{\'e}s-Alc{\'a}zar et al. in prep.]{Wellons2020,Parsotan2021,Cochrane2023}. 
Cosmological hyper-refinement simulations of this galaxy have shown explicitly that strong gravitational torques from stellar non-axisymmetries can drive large gas inflow rates down to sub-pc scales under these conditions \citep{Angles-Alcazar2021}, suggesting that this is a likely phase for strong AGN activity as well. We thus investigate the plausible positive and negative effects of AGN feedback at the galaxy's peak of nuclear and star formation activity. 

\subsection{Hyper-refined AGN winds} \label{subsec:BHdrivenwinds}

The method to inject AGN winds at super-Lagrangian resolution in cosmological simulations is described in Angl{\'e}s-Alc{\'a}zar et al. (in prep.), and builds on earlier particle spawning implementations in idealized simulations of galaxies and massive halos \citep{Richings2018,Torrey2020,Su2021}.
The BH is modelled as a collisionless particle with initial mass $M_{\rm BH}= 10^{9}\,{\rm M}_{\odot}$ and located near the centre of the main galaxy.
The accretion rate is assumed to be constant, for simplicity, throughout the duration of the simulation, representing a luminous quasar phase lasting $\sim 40\,{\rm Myr}$ with the BH accreting at a fixed fraction $\lambda_{\rm Edd}$ of the Eddington rate.
Stochastic swallowing of gas particles within the BH interaction kernel (defined to contain $\sim 256$ particles) ensures mass conservation \citep[e.g.][]{Angles-Alcazar2017a}. 

Our AGN wind model is specified by the following main properties: the mass outflow rate $\dot{M}_{\rm w}$, the initial wind velocity $v_{\rm w}$, and the geometry of the wind. We consider that a fraction $\epsilon_{\rm k}$ of the AGN bolometric luminosity ($L_{\rm bol} \equiv 0.1\,\dot{M}_{\rm BH}\,c^2$) emerges as a fast, nuclear isotropic wind radially outward from the BH, with initial velocity $v_{\rm w} = 30,000\,{\rm km}\,{\rm s}^{-1}$ and temperature $T_{\rm w}\sim 10^{4}\,{\rm K}$, typical of broad absorption line winds and ultrafast outflows \citep{Weymann1981,Gibson2009,Feruglio2015,Nardini2015,Tombesi2015}. We assume that the wind immediately interacts with the ambient medium, 
with post-shock velocity and temperature given by $v_{\rm sh} =v_{\rm w}/4 = 7,500\,\kms$ and $T_{\rm sh} \approx 1.2\times 10^{10}$\,K \citep{Faucher-Giguere2012}.
In practice, resolving the shock structure is challenging \citep{Richings2018,Richings2018-analytic,Torrey2020} and we model the AGN wind by creating or spawning new gas particles within a sphere $R_{\rm w}=0.1\,{\rm pc}$ around the BH, with initial velocity $v_{\rm sh}$ and temperature $T_{\rm sh}$ consistent with post-shock conditions. Other fluid quantities are immediately recomputed for the wind particles after spawning, interacting hydrodynamically with the ISM gas in the simulation. 
The simulations presented here implement a target wind particle mass of $1000\,\Msunh$, which represents a factor $>$20 times higher mass resolution than the original simulation, and we consider discrete ejection events containing between 10 and 100 wind particles distributed isotropically and moving radially outward from the BH.
Particle spawning allows us to fully capture the propagation and impact of fast winds at higher resolution than normally possible with Lagrangian hydrodynamics \citep[see also][]{Costa2020}, injecting feedback locally around the BH and capturing the wind-ISM interaction robustly regardless of gas geometry and at significantly higher resolution than nearest neighbor-based feedback coupling models. 

Table~\ref{table:models} summarizes the main properties of the simulations analysed here. All simulations start from the same initial conditions described in $\mathsection$\ref{subsec:initialconditions}, containing a central BH with mass $M_{\rm BH}=10^{9}\,{\rm M}_{\rm \odot}$, and implementing the same post-shock wind velocity and temperature while varying the mass outflow rate $\dot{M}_{\rm w}$. We assume that the BH accretes at the Eddington rate ($\lambda_{\rm Edd}=1$), motivated by hyper-refinement simulations that predict quasar-like inflow rates at sub-pc scales for the same simulated galaxy conditions \citep{Angles-Alcazar2021}. 
Along with the standard FIRE-2 simulation
that excludes AGN feedback (\nf), we investigate the impact of AGN-driven winds with kinetic feedback efficiencies in the range $\epsilon_{\rm k} =0.5$–50\%, which brackets a range of observational constraints \citep[e.g.][]{cicone2014,Fiore2017,harrison2018} and assumed feedback efficiencies in previous simulations \citep[e.g.][]{DiMatteo2005,Weinberger2017,Dave2019}.
Our choice in black hole mass and accretion rate is representative of those found in luminous quasars at $z\sim2$ given the host galaxy stellar mass ($\sim 10^{11}\,\Msun$; e.g., \citealt{Trakhtenbrot2014}; \citealt{Zakamska2019}). However, the assumed AGN wind kinetic efficiency exhibits a degeneracy with the chosen black hole mass and Eddington ratio. For example, by selecting a black hole mass or Eddington ratio a factor of 10 lower and simultaneously increasing our efficiencies by $\times$10, we would achieve equivalent mass, momentum, and energy loading factors for the AGN winds, which represent the actual relevant physical parameters in the simulations presented here.

The simulation name in each feedback implementation encodes the value of the mass loading factor ($\eta_{\rm k}$) and the kinetic feedback efficiency ($\epsilon_{\rm k}\times 100$).
The two simulations that we reference the most throughout this work are:
\begin{itemize}
    \item \nf: The control simulation using standard FIRE-2 physics, where we model the evolution of a massive galaxy ($M_{\rm star}\sim 10^{11}\,\Msun$) starting at $z\sim 2.28$ ($\Delta t =0$\,Myr) and no AGN winds are introduced. The BH is still accreting at the Eddington rate, $\dot{\rm M}_{\rm BH} \sim 22.2\,\Msunyr$.\\
    
    \item \mf: Fiducial simulation where AGN winds are turned on at $\Delta t =0$\,Myr with the same initial conditions as the \nf~case. We consider a luminous quasar phase with bolometric luminosity $L_{\rm bol} = 1.26\times 10^{47}$\,erg\,s$^{-1}$, driving a wind with kinetic efficiency $\epsilon_{\rm k}=0.1$ and mass loading factor $\eta_{\rm k}\equiv \dot{M}_{w}/\dot{M}_{\rm BH}= 2$, corresponding to a mass outflow rate in winds $\dot{M}_{w}=44.4\,\Msunyr$.
\end{itemize}
The times mentioned in this work are relative to the start of the wind phase at $t_{0}$, with $\Delta t$ referring to the time that has passed since then as $\Delta t\equiv t-t_{0}$.

\begin{figure*}
\includegraphics[width = \textwidth]{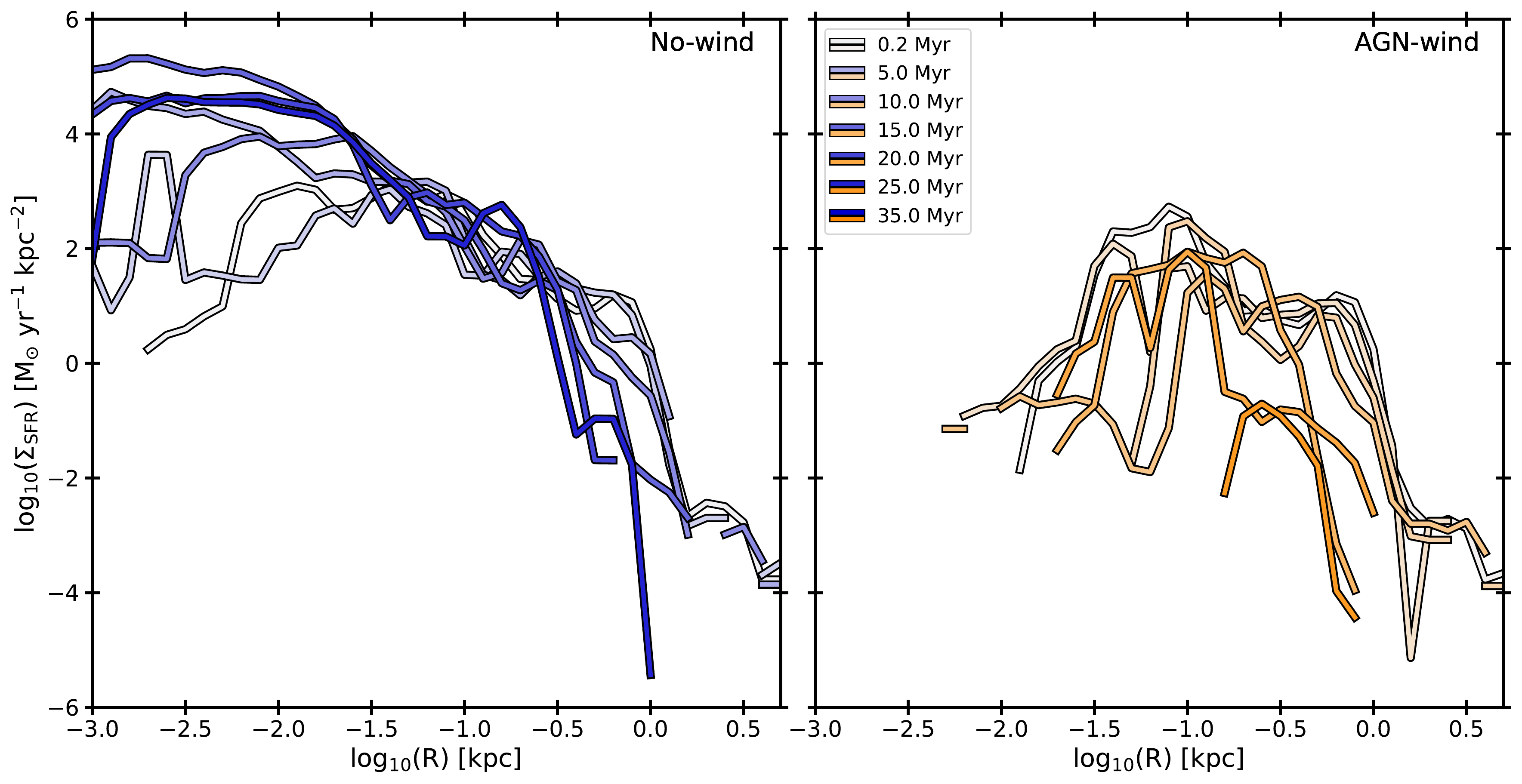}
\vspace*{-5mm}
\caption{\label{fig:RadialProfileall} Radial profile of star formation rate surface density ($\Sigma_{\rm SFR}$) for the \nf~(left; blue) and \mf~(right; orange) simulations. Time evolution is indicated by the saturation of the coloured lines. The high nuclear $\Sigma_{\rm SFR}$ in the \nf~simulation continues to increase with time while AGN winds create a central cavity and suppress the overall SFR.}
\end{figure*}

\section{Overview of Simulations} \label{Overview of Simulations}

Figure~\ref{fig:SFRall} shows the projected star formation rate surface density for two different simulations of the same massive star-forming galaxy ($M_{\rm star}\sim10^{11}\,{\rm M}_{\odot}$, ${\rm SFR}\sim300\, {\rm M}_{\odot}\, {\rm yr}^{-1}$) at $z\sim2.28$, one with no AGN feedback (\nf), and the other with feedback (\mf), at varying snapshots in time. In the \nf~case (top two rows), the central BH accretes gas at the Eddington limit but we neglect any AGN feedback effects, with stellar feedback solely responsible for regulating star formation. At the beginning of the simulated period, the galaxy resembles a turbulent, clumpy, kpc-scale disc with significant amounts of star formation occurring along fractured spiral arms and the denser nuclear region. As time proceeds, the dense, star-forming gas reservoir continues to be replenished by infalling gas across scales and becomes more concentrated toward the centre, forming an ultra-compact, nuclear star-forming disc after $\Delta t \sim 25$\,Myr of evolution. From the first snapshot ($\Delta t = 0.2$\,Myr) to the last ($\Delta t = 35$\,Myr), the SFR in the central 1\,kpc region increases from $\sim 400\, {\rm M}_{\odot}$ ${\rm yr}^{-1}$ to $\sim 640\, {\rm M}_{\odot}$ ${\rm yr}^{-1}$.

In the \mf~simulation (bottom two rows), we inject radial shells of high resolution particles to represent the AGN wind. In the first $\Delta t = 1$\,Myr of evolution there is already a noticeable effect on the galaxy. The wind has ejected star-forming gas from the central 50\,pc as well as cleared out other areas throughout the disc that had high SFR (dark green regions) in the \nf~case, that now have lower amounts of star formation (lighter green regions) or none at all. Winds propagate more efficiently along paths of least resistance, preferentially along the rotation axis of the galaxy, but can nonetheless penetrate through low-density ISM channels across the galaxy disc.  After 5\,Myr, the morphology of the star-forming gas becomes drastically different, with AGN winds opening up a growing, central cavity surrounded by strips of gas, in contrast to the ultra-compact star-forming disc in the absence of AGN feedback. Not only is the morphology different than in the \nf~case, but the amount of star formation is drastically different as well. By $\Delta t = 0.2$\,Myr, the AGN winds have already had a clear impact on the nuclear SFR, with $\sim230\,{\rm M}_{\odot}\,{\rm yr}^{-1}$ within 1\,kpc, a factor of $\sim0.5$ less than in the \nf~simulation. At a later time, 25 Myr, the total SFR in the \mf~case reaches $\sim 19\,{\rm M}_{\odot}\,{\rm yr}^{-1}$, constituting a decrease by a factor of $\sim33$ from the \nf~case, an overwhelmingly negative effect. By the end of the simulation, ($\Delta t=35$\,Myr), the total SFR within 1\,kpc in the \nf~case is $\sim 640\,{\rm M}_{\odot}\,{\rm yr}^{-1}$, but only $\sim 0.4\,\Msunyr$ in the \mf~case, where most of the dense star-forming gas has been evacuated by the winds. In the central 250\,pc region, the SFR increases from $\sim 230\, {\rm M}_{\odot}$ ${\rm yr}^{-1}$ ($\Delta t =0.2$\,Myr) to $\sim 600\, {\rm M}_{\odot}$ ${\rm yr}^{-1}$ ($\Delta t =35$\,Myr) in the \nf~simulation, while in the \mf~case it goes from $\sim 90\, {\rm M}_{\odot}$ ${\rm yr}^{-1}$ ($\Delta t =0.2$\,Myr) to $\sim 0.003\, {\rm M}_{\odot}$ ${\rm yr}^{-1}$ ($\Delta t =35$\,Myr). While the global SFR increases by 40\% during the \nf~simulation, the SFR in the \mf~case decreases by a factor of $\sim 3000$.

Figure~\ref{fig:RVvsRDall} shows the 2D distribution of star-forming gas in the radial velocity--radial distance plane for the \nf~(top) and \mf~(bottom) simulations. For each case, we show separately the inflowing ($v_{\rm r} < 0$) and outflowing ($v_{\rm r} > 0$) components of the star-forming gas, which are defined relative to the radial velocity ($v_{\rm r}$) with respect to the BH. We thus refer to inflows and outflows as radially inward and outward components without requiring a minimum absolute velocity.
Some of the morphological features that were present in Figure~\ref{fig:SFRall} are identifiable within the 2D distribution for the \nf~simulation, such as the vertical stripes corresponding to spiral arms. As time progresses, the star-forming gas becomes more centrally concentrated,
which is seen here as the distribution shifting to the left.  Aside from the increase in total SFR, there is an increase in the net radial velocity with time, reaching $\sim$2,000 km/s on 10--100\,pc scales for both the outflowing and inflowing components ($\Delta t=20-25\,{\rm Myr}$). In the \nf~simulation, the two radial velocity components show similar distributions and contribute approximately similar amounts to the global SFR, with a slight predominance of the inflow component at later times. In the absence of coherent, large scale inflow/outflow conditions dominating the gas' dynamics, the roughly symmetric radial velocity structure is the result of highly turbulent motions in the ISM.  The intense star formation in the nuclear region at high surface densities ($\Sigma_{\rm SFR} > 1000\,\Msunyr\,{\rm kpc}^{-2}$) rapidly deepens the potential well, resulting in the increased maximum radial velocities observed.

In the \mf~simulation, the radial distribution of star-forming gas clearly shows the impact of AGN feedback, with the size of the central cavity growing with time on average (bottom rows).  Despite the injection of positive radial momentum, there is no significant sign of SFR enhancement in the outflowing component. During the first $\sim$10\,Myr of evolution, the inflow and outflow components contribute roughly equally to the total SFR, similar to the \nf~case. However, the inflow component tends to dominate at later times, when the impact of winds becomes more dramatic. This suggests that gas radially accelerated by the winds has generally lower probability of forming stars. One interesting exception is the outflowing component at $\Delta t = 25$\,Myr since the start of the AGN wind phase, which contains a considerable amount of star formation (16\,$\Msunyr$, corresponding to $>$80\%~of the total SFR) in a localized region at 300\,pc from the centre and with radial velocity as high as 800\,${\rm km}\,{\rm s}^{-1}$ on average.

\section{Negative Global Impact of AGN feedback} \label{Negative Impact of AGN feedback}

Figure~\ref{fig:RadialProfileall} shows the azimuthally-averaged radial profile of the star formation rate surface density for the \nf~(left) and \mf~(right) simulations, for various times. We use cylindrical radial bins that are defined relative to the angular momentum rotation axis of star-forming gas within 2\,kpc (since the majority of the star-forming gas is contained within this scale) of width $\Delta\log_{10}(R) = 0.1\,{\rm dex}$ to calculate the SFR per unit area in each bin ($\Sigma_{\rm SFR}$). For clarity, we smooth the resulting radial distributions by applying a running average considering the two nearest neighbor bins. At the start of the simulation in the \nf~case, the star-forming disc reaches $\Sigma_{\rm SFR} \sim 1000\,\Msunyr\,{\rm kpc}^{-2}$~in the nuclear region (10--100\,pc), while maintaining $\Sigma_{\rm SFR} \sim 10\,\Msunyr\,{\rm kpc}^{-2}$~on kpc scales.  In the absence of AGN winds, the nuclear gravitational potential deepens significantly and the radial distribution of star-forming gas steepens strongly over the 35\,Myr period, reaching extreme surface densities, up to $\Sigma_{\rm SFR} \sim 10^4\,\Msunyr\,{\rm kpc}^{-2}$ in the inner pc, and containing most star formation within $\sim$200\,pc.

In contrast to the \nf~simulation, the formation of the central cavity is clearly visible in the radial SFR profile of the \mf~simulation. The size of the cavity increases with time, on average, as winds continue to inject energy and momentum into the surrounding gas. However, the persistent, non-isotropic infall of gas onto the galaxy and the inefficient coupling of winds with low-subtended area gas structures drives fluctuations in the size of the cavity, with dense gas clumps sometimes able to penetrate the inner cavity down to $<$10\,pc scales. Besides the cavity opening, the amplitude of the $\Sigma_{\rm SFR}$ profile also decreases with time across radial scales, emphasizing the overall negative impact of AGN winds on the star formation properties of the host galaxy.

\begin{figure}
\includegraphics[width = \columnwidth ,valign=t]{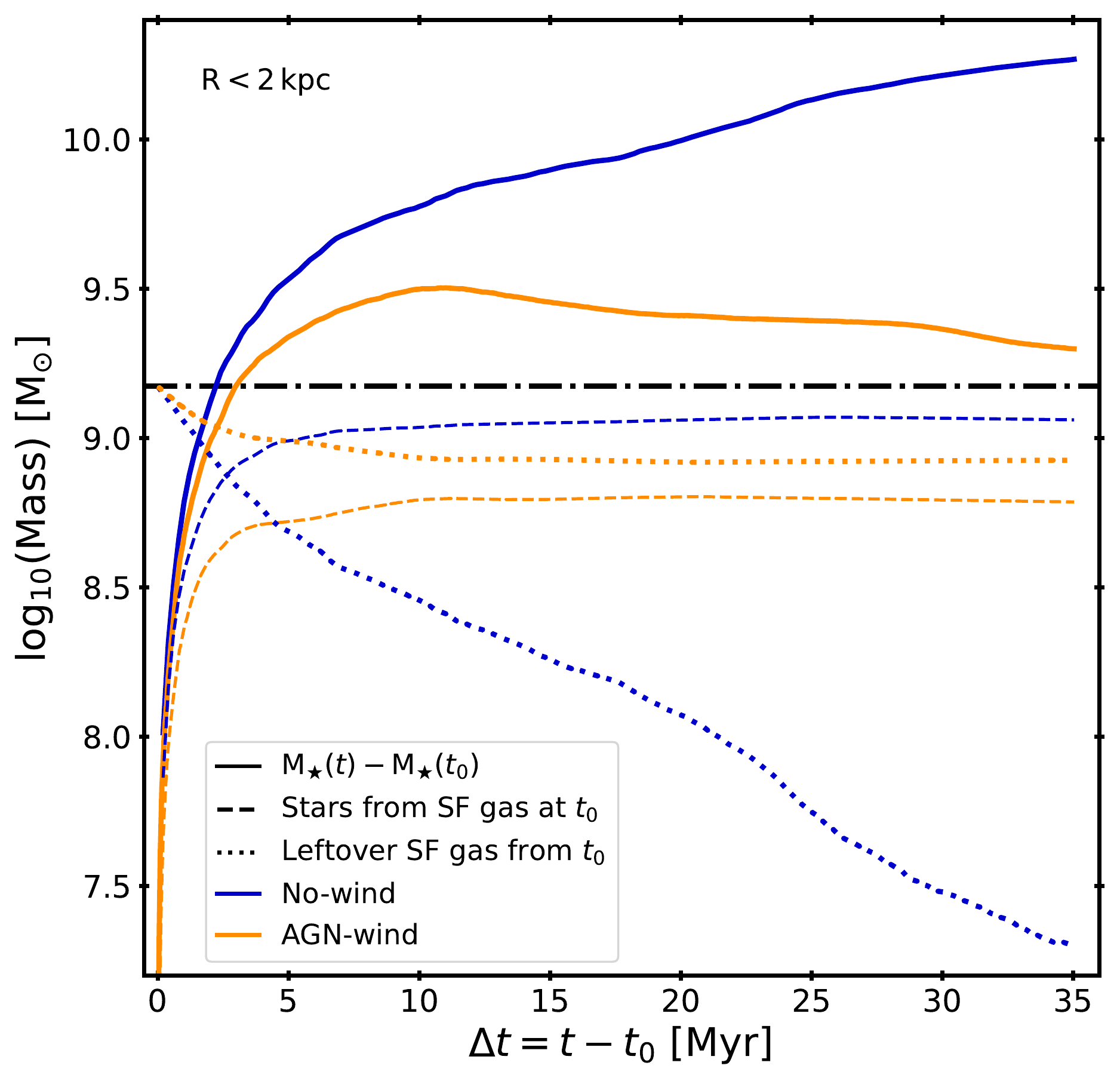}
\vspace*{-3mm}
\caption{\label{fig:stellarmass} Stellar mass growth within the central 2\,kpc as a function of time for the \nf~(blue) and \mf~(orange) simulations. Solid lines represent the total mass of stars formed since $t_{0}$, i.e. excluding any pre-existing stars. The horizontal dash-dotted line (black) represents the initial total mass of star-forming gas available at the start of the simulations ($t_{0}$), the dashed lines indicate the stellar mass formed from this gas, and the dotted lines show the remaining gas mass from originally star-forming gas. The suppression of stellar mass growth by AGN winds over $\sim$35\,Myr is driven primarily by a reduction in the amount of new gas that can become star-forming as opposed to by directly ejecting pre-existing star-forming gas.}
\end{figure}

Figure~\ref{fig:stellarmass} shows the total stellar mass enclosed in the central 2\,kpc of the galaxy as a function of time since the start of the AGN feedback phase at $\Delta t = 0$, excluding stars formed at earlier times (solid lines). 
In the \nf~simulation (blue), the extreme SFR surface densities reached lead to the formation of $\sim 2\times 10^{10}\,\Msun$~of stars during only 35\,Myr. In contrast, the overall reduction in SFR in the AGN-wind simulation (orange) yields the formation of only $\sim 3\times 10^{9}\,\Msun$~of stars during the initial 10\,Myr, with a subsequent $\sim$60\% decrease in enclosed stellar mass within 2\,kpc owing to the expansion of the stellar component driven by the expulsion of the nuclear gas reservoir as well as stellar mass return to the ISM and significantly lower SFR at later times.
The effect of stellar migration, included in this analysis, is a less direct and yet important mechanism by which AGN winds impact the stellar sizes of galaxies \citep{Cochrane2023}.

In order to understand the effect of AGN winds, we identify all gas elements within 2\,kpc that have non-zero SFR at $t=t_{0}$ and track them in time by means of their unique identifiers, which are preserved when gas elements are converted into star particles \citep[e.g.][]{Angles-Alcazar2017b}. The horizontal dash-dotted line (black) indicates the corresponding total initial amount of star-forming gas available at $t=t_{0}$. The blue and orange dashed lines represent the amount of stellar mass formed from this selected initial star-forming gas reservoir in the \nf~and \mf~simulations, respectively, while the leftover gas mass is shown as the dotted lines.

In the \nf~simulation, almost all of the initial star-forming gas is converted into stars after 35\,Myr, corresponding to $\sim 10^9\,\Msun$\footnote{There is $\sim$20\%~of ``missing'' mass due to stellar mass loss as well as BH accretion.}. In contrast, in the \mf~simulation, only $\sim$40\%~of the original star-forming gas is converted into stars ($\sim6\times 10^{8}\,\Msun$), which demonstrates a direct, negative impact of AGN winds on the pre-existing star-forming gas. However, the total amount of stellar mass formed in the \nf~case is more than one order of magnitude larger than the mass available in the initial star-forming gas reservoir, indicating that the majority of stars form from additional gas becoming star-forming over time. The dominant negative effect of AGN winds over the 35\,Myr period is thus to prevent the replenishment of the star-forming gas reservoir, with the direct ejection of pre-existing star-forming gas playing a lesser role.

\begin{figure*}
\centerline{\includegraphics[scale=.5,valign=t]{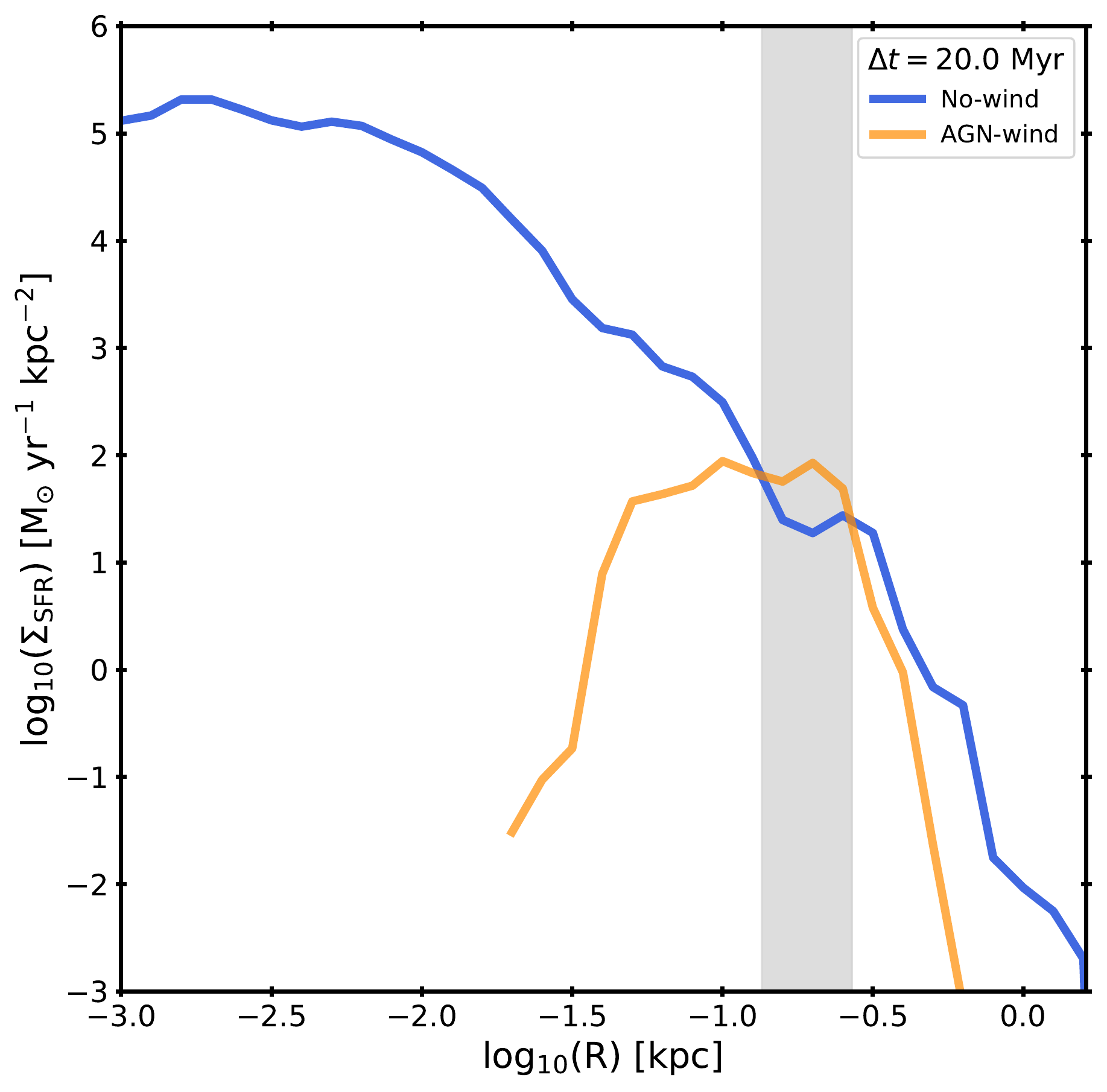} \includegraphics[scale=.5,valign=t]{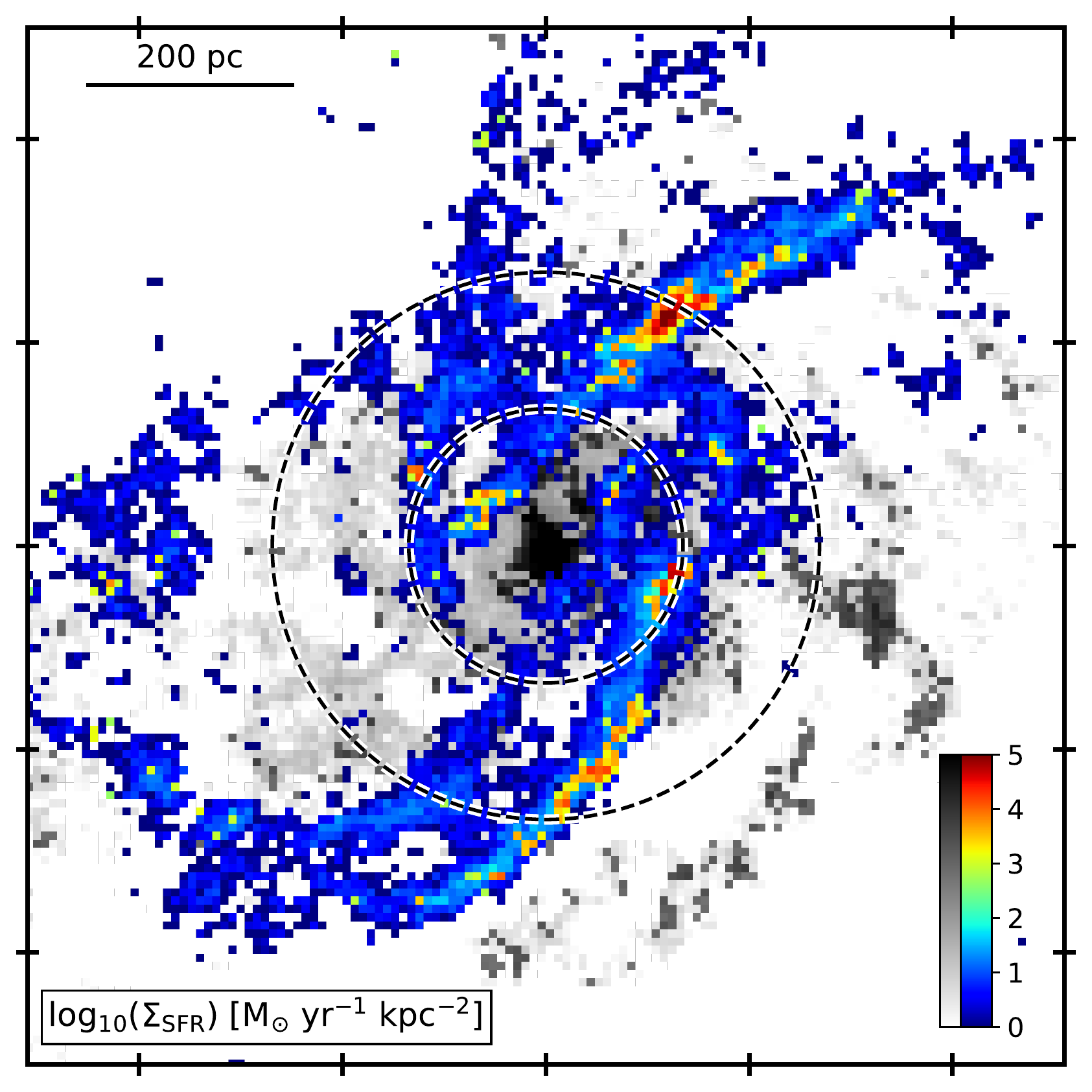}}
\vspace*{-2mm}
\caption{\label{fig:mildfeedback_high} Left: Radial profile of the total star formation rate surface density ($\Sigma_{\rm SFR}$) corresponding to $\Delta t = 20$\,Myr since the start of the AGN feedback phase. The blue line represents the \nf~case while orange is the \mf~case. The vertical grey band indicates the region where $\Sigma_{\rm SFR}$ is higher in the \mf~simulation compared to the \nf~case. Right: Projected star formation surface density map with the \nf~(grey scale) and \mf~(colour scale) simulations superimposed. The radial annulus corresponds to the vertical grey band in the left panel. Despite the overall negative effect of BH feedback, $\Sigma_{\rm SFR}$ can reach higher values in certain regions under the presence of AGN winds, suggesting local positive feedback effects.}
\end{figure*}

\section{Signatures of (Local) Positive AGN feedback} \label{Signatures of (Local) Positive AGN feedback}
\subsection{Spatial anti-correlation of winds and star-forming regions} \label{Spatial anti-correlation of winds and star-forming regions}

Figure~\ref{fig:mildfeedback_high} shows the star formation rate surface density as a function of radial distance (left panel), comparing the \nf~and \mf~simulations at time $\Delta t = 20$\,Myr. This highlights again the overall negative impact of AGN winds, creating a central cavity devoid of star-forming gas, in stark contrast to the extreme $\Sigma_{\rm SFR}$ values reached in the absence of winds. However, this also shows that $\Sigma_{\rm SFR}$ can be larger in the presence of AGN winds in localized regions under certain conditions. In this case, the azimuthally-averaged $\Sigma_{\rm SFR}$ at a radial distance of $\sim$150--250\,pc is larger in the AGN-wind simulation compared to the \nf~simulation (grey shaded area), which represents a plausible indication of local positive AGN feedback.

The same effect is illustrated in the right panel, where we show the projected, face-on spatial distribution of star-forming gas, overlaying the \nf~(grey scale) and \mf~(colour scale) simulations.  Most star formation in the presence of AGN winds occurs in dense gas clumps located at $>$100\,pc, outside of the wind-dominated region. This spatial anti-correlation of winds and high star-forming regions suggests that the accumulation and compression of gas by AGN feedback in certain regions can {\it enhance}, rather than suppress, local star formation.

\subsection{Enhanced star formation efficiency by AGN wind compression} \label{Enhanced star formation efficiency by AGN wind compression}

\begin{figure*}
\centerline{\includegraphics[scale=.465,valign=t]{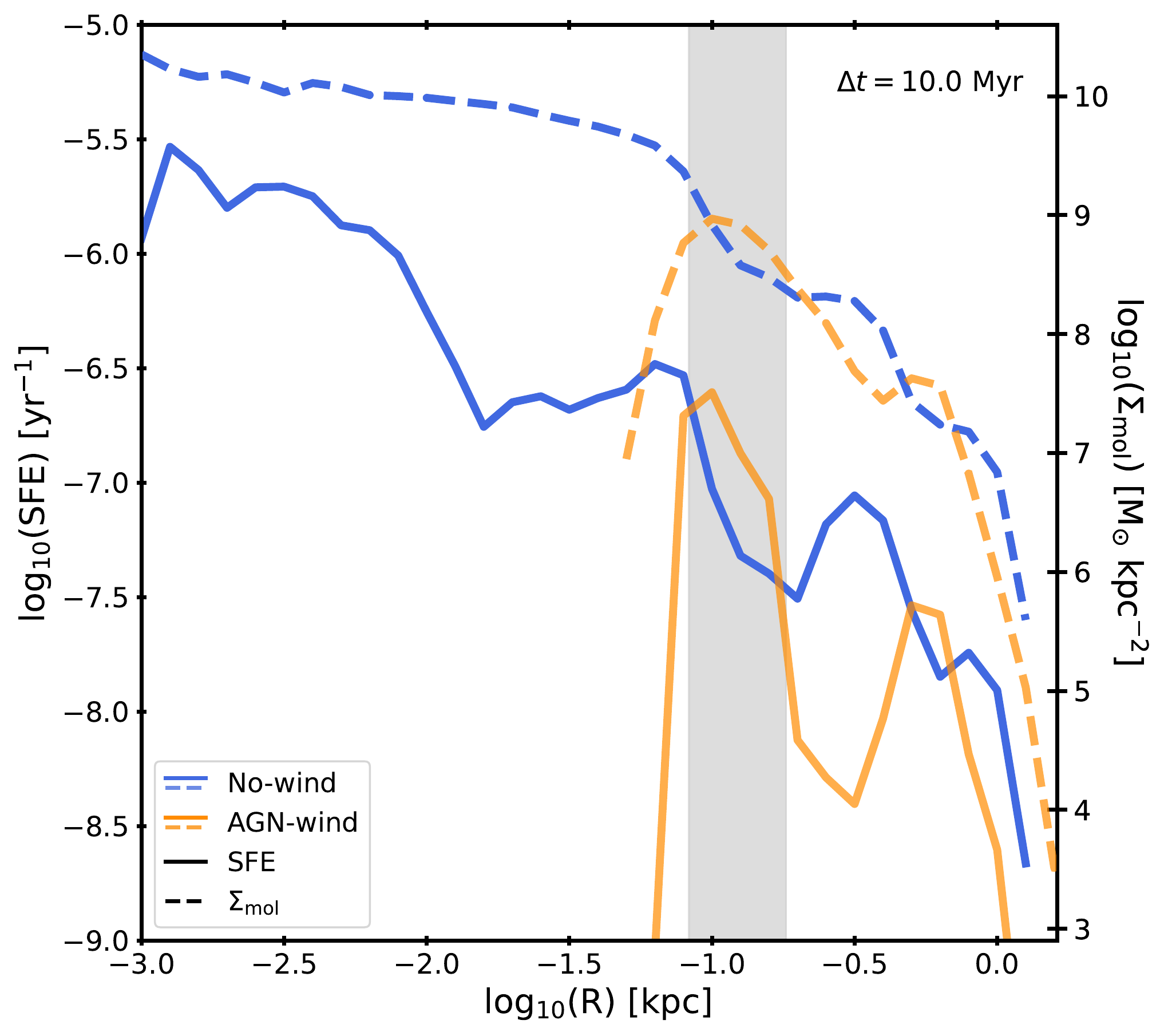} \includegraphics[scale=.47,valign=t]{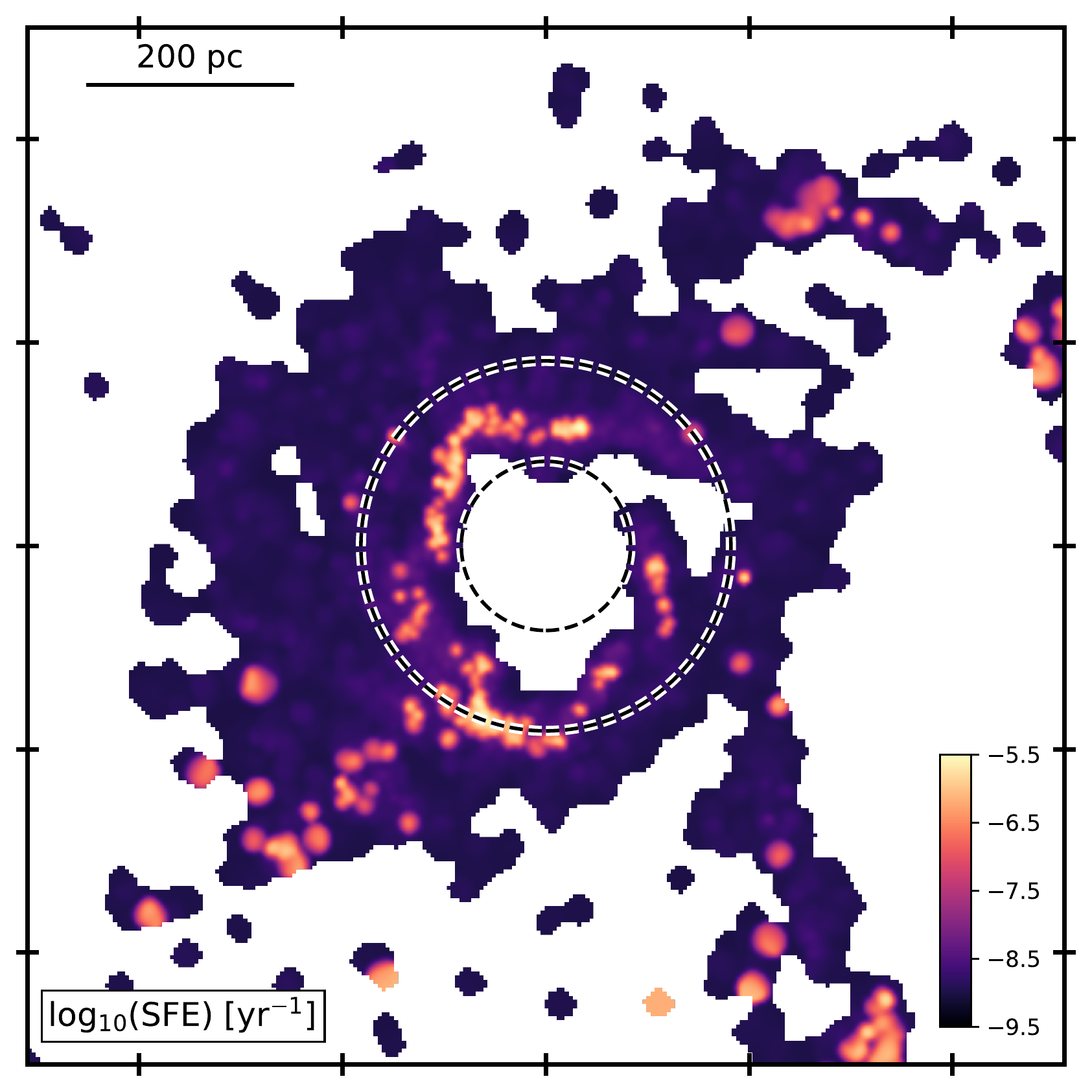}}
\centerline{\includegraphics[scale=.465,valign=t]{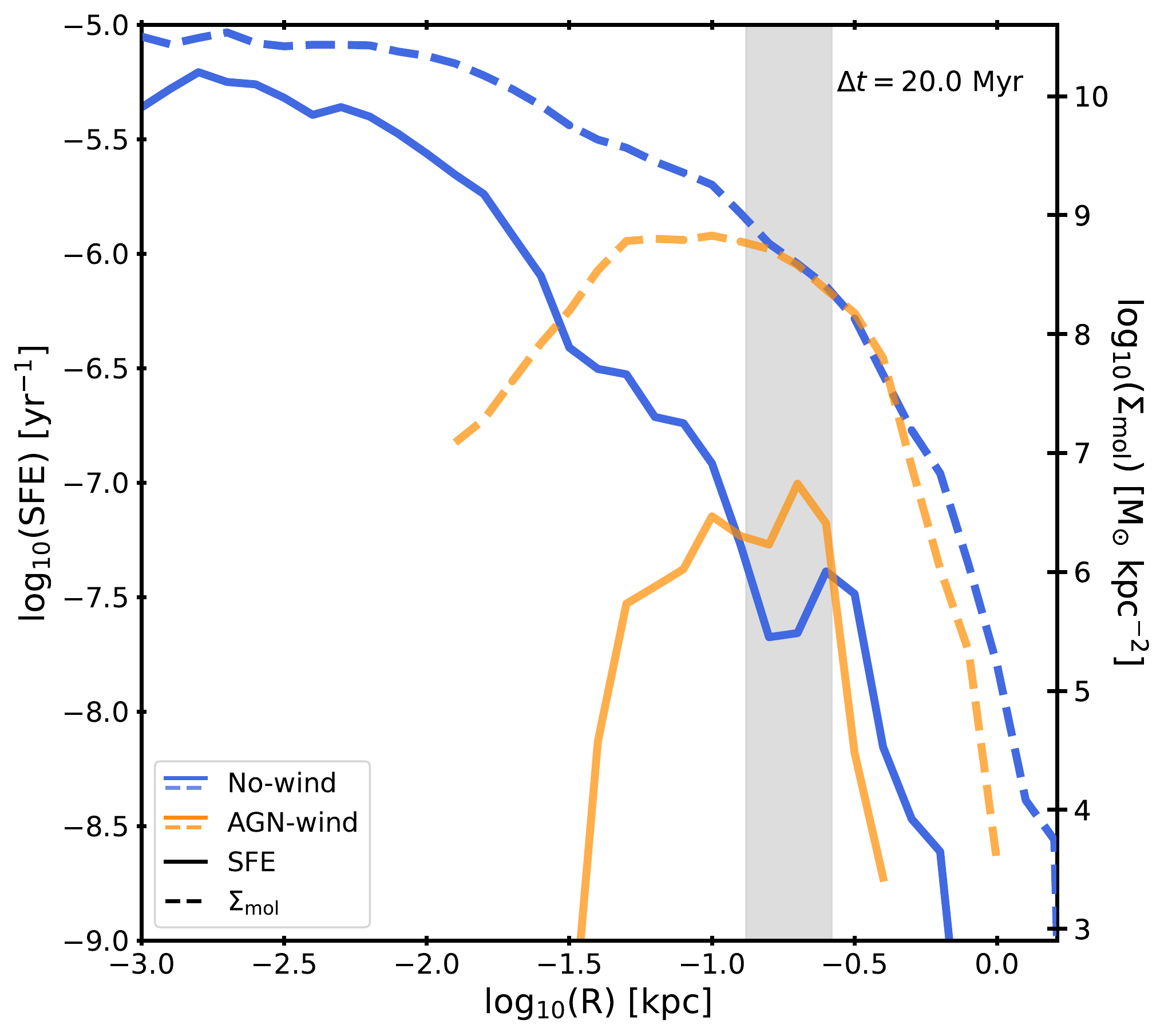} \includegraphics[scale=.47,valign=t]{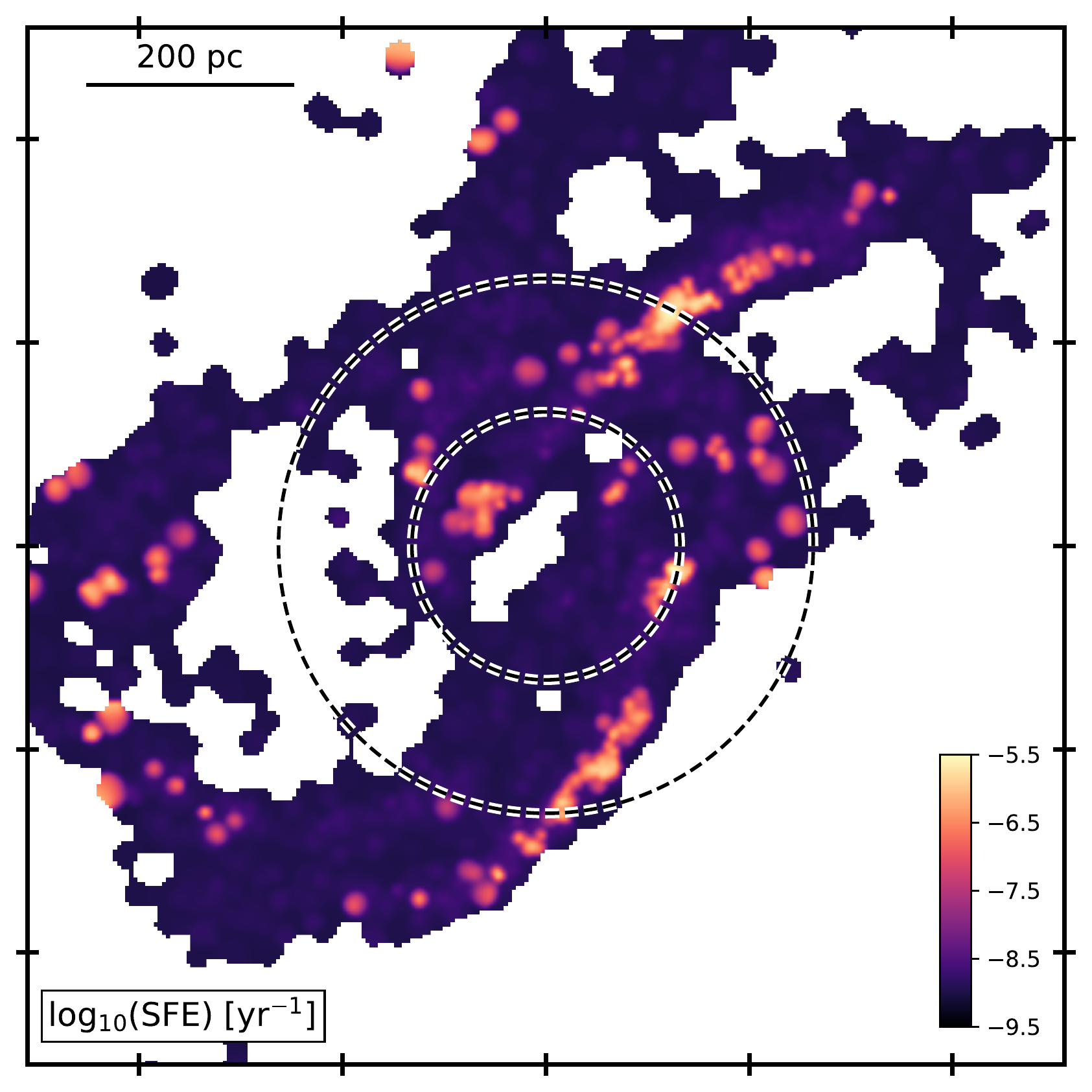}}
\vspace*{-2mm}
\caption{\label{fig:mildfeedback_sfehigh_mgm} Left:  Radial profile of the star formation efficiency (SFE; solid lines) and the total molecular gas surface density ($\Sigma_{\rm mol}$; dashed lines) at time $\Delta t = 10$\,Myr (top) and $\Delta t = 20$\,Myr (bottom) since the beginning of the AGN feedback phase for the \nf~(blue) and \mf~(orange) simulations.  The vertical grey band indicates the region where SFE is higher in the \mf~simulation compared to the \nf~case. Right: Spatial SFE distribution for a face-on projection of the \mf~case. The radial annulus corresponds to the vertical grey band in the left panel. The star formation efficiency increases along the edge of the central cavity relative to the \nf~simulation, suggesting that compression of ISM by AGN winds can trigger star formation locally.}
\end{figure*}

Figure~\ref{fig:mildfeedback_sfehigh_mgm} investigates in more detail the local triggering of star formation by AGN winds, comparing the radial profiles of star formation efficiency (SFE) and molecular gas mass surface density ($\Sigma_{\rm mol}$) at times $\Delta t = 10$\,Myr and $\Delta t = 20$\,Myr for the \nf~and \mf~simulations (left panels). We compute the SFE of each gas element as the SFR divided by mass, ${\rm SFE}\equiv{\rm SFR}/{\rm M}_{\rm gas}$, which is then averaged (mass-weighted) over all gas elements in cylindrical radial bins. For $\Sigma_{\rm mol}$, we assume that gas elements with Hydrogen number density $n_{\rm H}>1000$\,cm$^{-3}$ have a molecular fraction near unity. We also show face-on projections of the spatial distribution of SFE for the \mf~simulation (right panels).

In both simulations we see a consistent trend of higher SFE with higher molecular gas surface density, as expected, but they do not follow a one-to-one relation because the SFR of a given gas clump depends on its free-fall time, virial parameter, and other factors \citep{Hopkins2018}. In the \nf~simulation, SFE and $\Sigma_{\rm mol}$ both increase by more than three orders of magnitude from kpc to pc scales at $\Delta t = 10$\,Myr, and by almost five orders of magnitude at $\Delta t=20$\,Myr, since gas had more time to accumulate in the centre, becoming more centrally concentrated.
In the presence of AGN feedback, SFE and $\Sigma_{\rm mol}$ also tend to increase with decreasing radial distance, but both drop very rapidly inside of the central cavity evacuated by the AGN winds.

A local enhancement of SFR in the presence of AGN winds relative to the \nf~case, such as seen at $\Delta t = 20$\,Myr (Figure~\ref{fig:mildfeedback_high}), could be due to an increase in the amount of dense gas available in a given region and/or an increase in the efficiency of converting gas into stars \citep{Moreno2021}. The bottom left panel of Figure~\ref{fig:mildfeedback_sfehigh_mgm} shows that, at $\Delta t = 20$\,Myr, the increase in local SFR surface density in the range $\sim$150--250\,pc under the presence of AGN winds coincides with the SFE being up to one order of magnitude higher in the \mf~simulation compared to the \nf~case. In the same radial range, however, the amount of fuel for star formation (molecular gas) is similar both in the presence and absence of AGN winds, suggesting that the increase in local SFR is more efficiency-driven than fuel-driven in this case. We find similar results when considering the SFE per free-fall time (Fig.~\ref{fig:mildfeedback_sfehigh_mgm}).

The earlier physical conditions at $\Delta t = 10$\,Myr (Figure~\ref{fig:mildfeedback_sfehigh_mgm}, top panels) also represent an interesting example of increased SFE locally ($R \sim 100$\,pc) in the presence of AGN winds despite the overall suppression of global star formation relative to the \nf~simulation. The top right panel of Fig \ref{fig:mildfeedback_sfehigh_mgm} shows that gas clumps with high SFE clearly form along the edge of the cavity, where the incident AGN winds interact with the ISM. In this case, they also have higher molecular gas surface density compared to the \nf~simulation. 
AGN winds can be an efficient mechanism to redistribute metals from the center of the galaxy to larger scales \citep[e.g.][]{Sanchez2019,Choi2020,Sanchez2023}, and metal rich gas pushed out from the center could perhaps increase the cooling rate of gas along the edge of the cavity and contribute to the higher SFE that we see. However, this is likely a subdominant effect given that the metallicity is significantly higher in the \nf~simulation throughout the entire central 1\,kpc region, which could not explain the higher local SFE seen in the \mf~case along the edge of the cavity. There are indeed no signs of increased metallicity in the \mf~simulation, which can be explained by the increasing amount of gas ejected in winds along the polar direction as the cavity radius increases \citep{Torrey2020}, depositing metals preferentially outside of the galaxy rather than increasing the metallicity of the remaining ISM gas.

Overall, our results suggest that AGN winds can trigger star formation and enhance the local star formation efficiency locally by compressing the ISM, while having a global negative effect. 

\subsection{Contribution of outflowing gas to star formation} \label{Contribution of outflowing gas to star formation}
\begin{figure*}
\centerline{\includegraphics[scale=.5,valign=t]{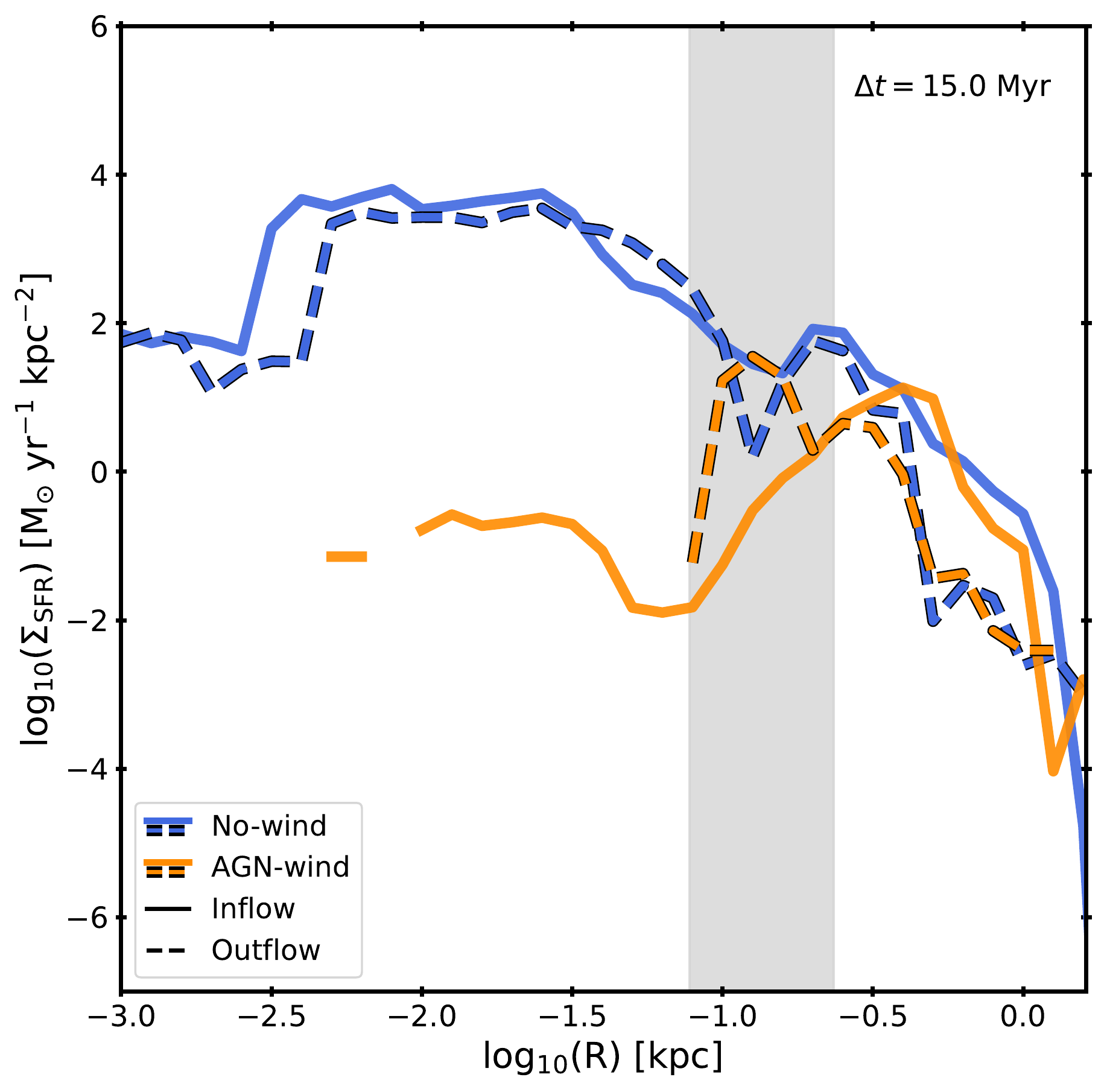} \includegraphics[scale=.5,valign=t]{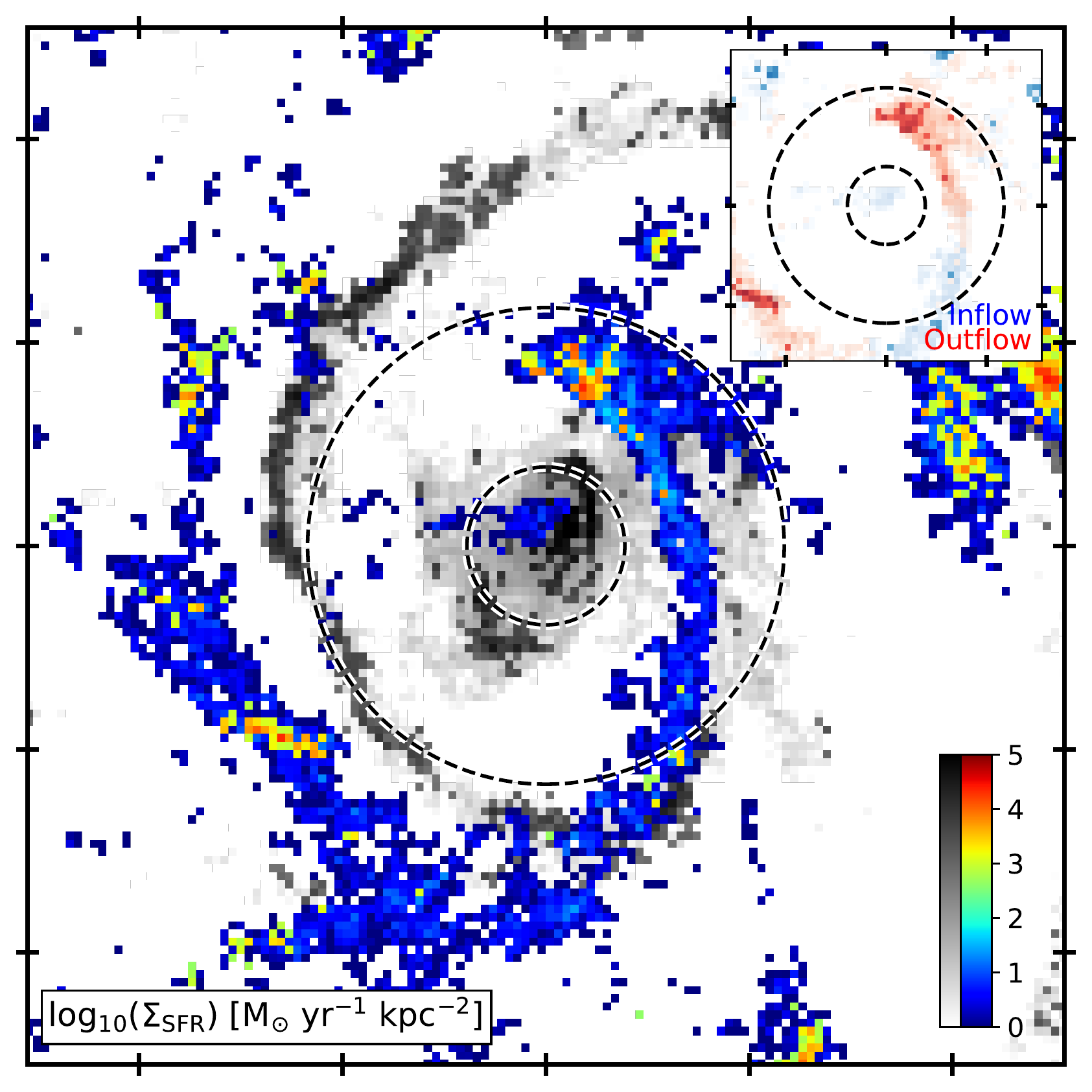}}
\centerline{\includegraphics[scale=.5,valign=t]{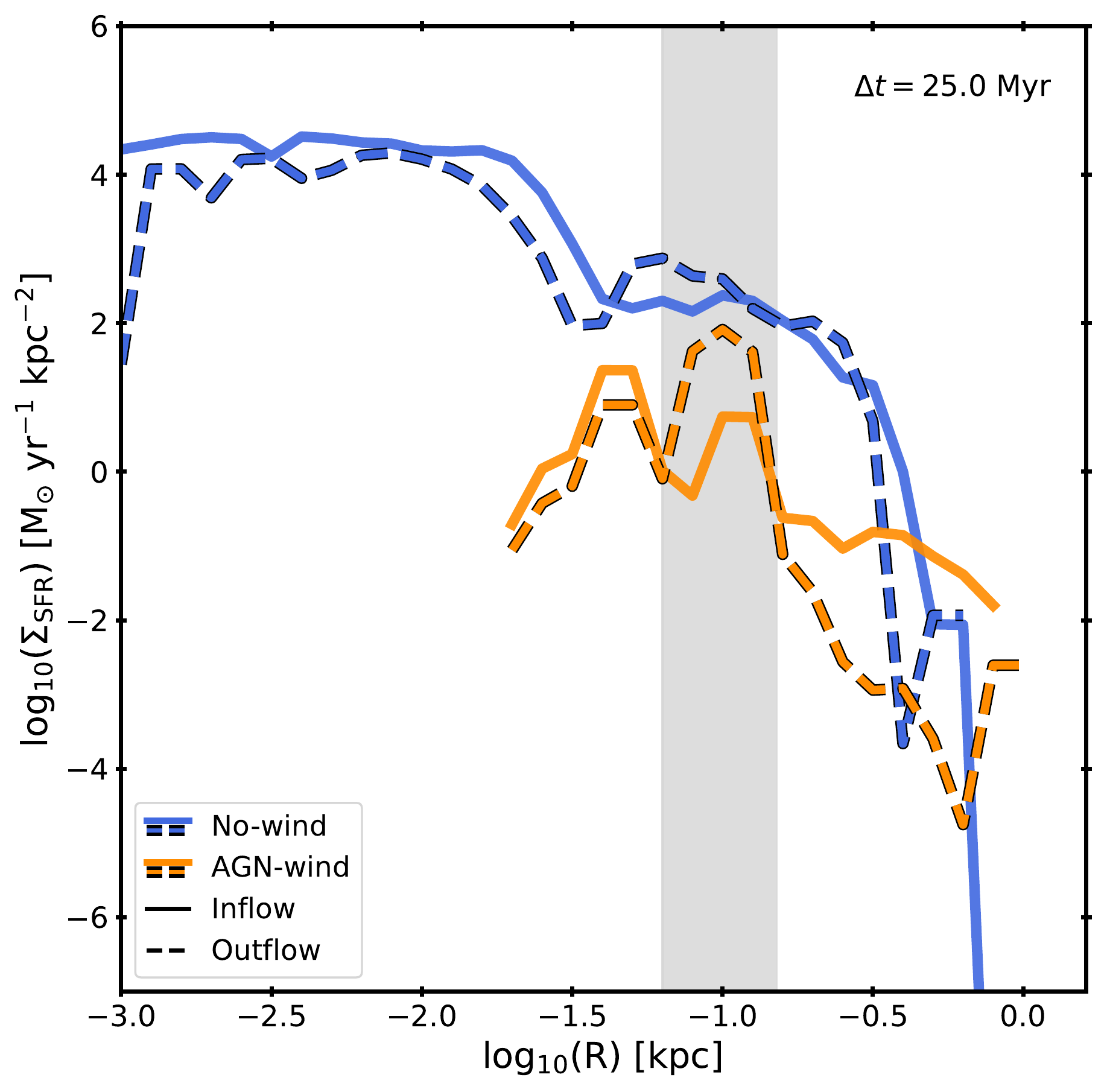} \includegraphics[scale=.5,valign=t]{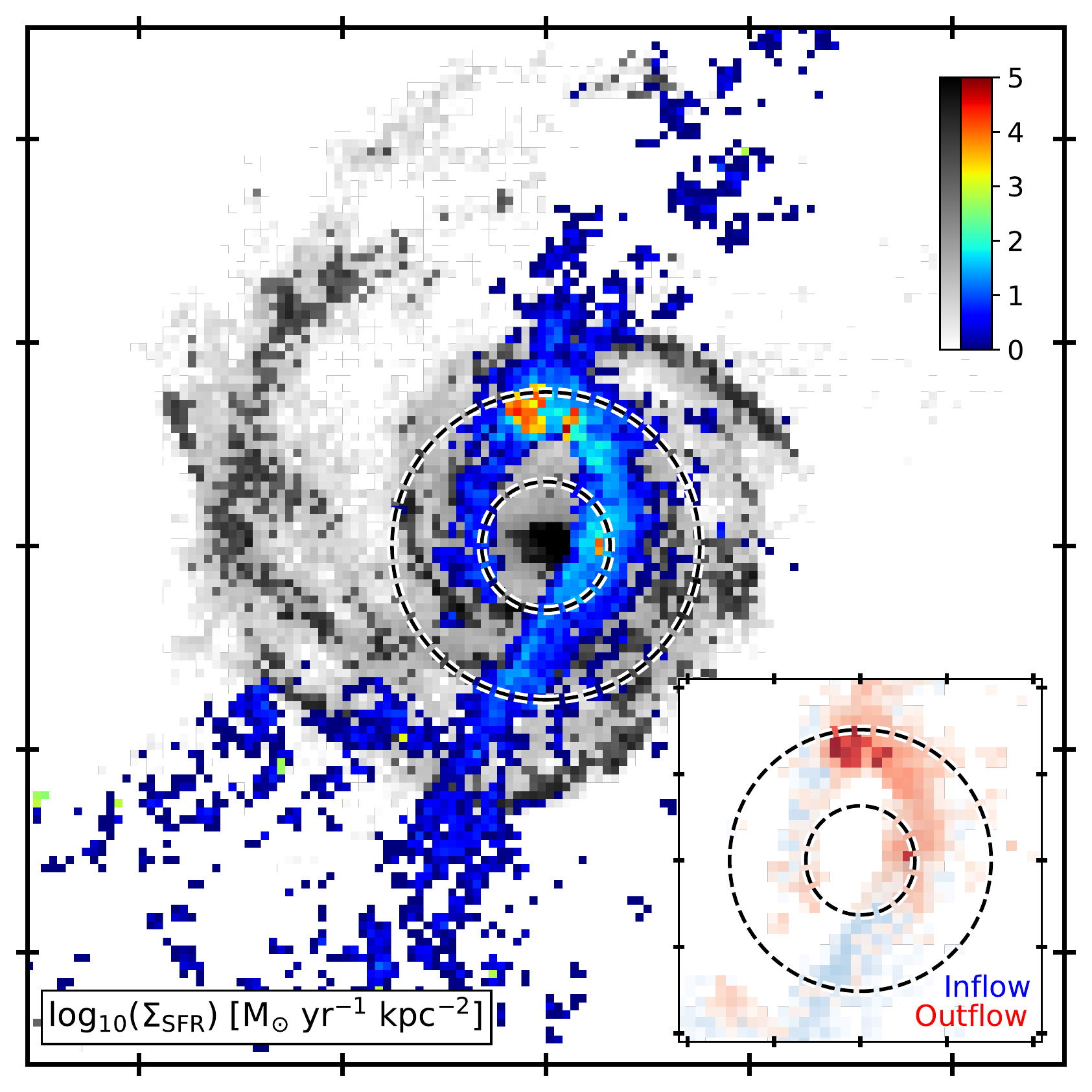}}
\vspace*{-2mm}
\caption{\label{fig:outflowhigh} Left: Radial profile of the total star formation rate surface density (\sigmasfr) for the inflowing material (solid lines) and outflowing material (dashed lines) at time $\Delta t = 15$\,Myr (top) and $\Delta t = 25$\,Myr (bottom) since the beginning of the AGN feedback phase for the \nf~(blue) and \mf~(orange) simulations. Right: Spatial star formation surface density distribution for a face-on projection with the \nf~(grey scale) and \mf~(colour scale) simulations superimposed. Inset: A closer look at the central 250\,pc (top) and 200\,pc (bottom) region, decomposing the \mf~map into inflowing (blue) and outflowing (red) components. The vertical grey band (left) and corresponding radial annulus (right) indicate the region where the outflowing component dominates over the inflowing component for the \mf~simulation. The increased local fractional contribution of outflowing gas to star formation is a plausible signature of positive AGN feedback.}
\end{figure*}

Figure~\ref{fig:outflowhigh} shows the radial profiles of the SFR surface density at $\Delta t=15$\,Myr and $\Delta t=25$\,Myr (left), separating the distribution into inflowing and outflowing components based on the radial velocity of gas, as in Figure~\ref{fig:RVvsRDall}. We also show the projected, face-on distributions of $\Sigma_{\rm SFR}$ in the right panels, overlaying the \nf~(grey scale) and \mf~(colour scale) simulations.

In the \nf~simulation at $\Delta t = 15$\,Myr, a single spiral arm surrounds the ultra-compact nuclear disc forming on 100\,pc scales. As shown above, introducing AGN feedback changes the morphology of the galaxy, forming a slightly off-centred cavity $\sim$400\,pc wide by evacuating the nuclear gas disc and pushing outward the dominant spiral arm.  The inset figure in the $\Sigma_{\rm SFR}$ map indicates the spatial distribution of the inflowing (blue) and outflowing (red) star-forming gas, showing that the region near the cavity edge is dominated by the outflow component. Interestingly, the inflow component reaches inside of the cavity down to the inner $\sim$10\,pc, previously devoid of dense gas, despite the continuous injection of winds pushing material outwards. At $\Delta t = 25$\,Myr, the centre of the cavity has been cleared out again by the winds, but the overall size of the cavity has decreased to $\sim$100\,pc in diameter, owing to the large inflow of gas occurring during the previous $\sim$10\,Myr.  At this time, the dense, ring-like gas structure in the \mf~simulation is outflow-dominated and on the way to expand again, in contrast to the ultra-compact nuclear spiral prevalent in the \nf~simulation. This illustrates the complex interaction between AGN-driven winds and infalling gas in the strong nuclear gravitational potential of a massive galaxy. 

In the \nf~simulation, the inflow and outflow components of star-forming gas exhibit roughly similar radial distributions, indicative of turbulent motions (see also Figure~\ref{fig:RVvsRDall}), with the overall contribution of inflowing material to global star formation being slightly predominant over the outflow component for both $\Delta t = 15$\,Myr and $\Delta t = 25$\,Myr. In the \mf~simulation, however, we identify a decoupling of dynamical components, with certain regions clearly dominated by outflowing material. In the radial range indicated by the vertical grey bands, the outflow component reaches $\Sigma_{\rm SFR}$ values between one ($\Delta t = 25$\,Myr) and two ($\Delta t = 15$\,Myr) orders of magnitude higher than the inflow component, coinciding with gas structures along the edge of the cavity compressed by the AGN winds. Thus, even in cases such as these, where AGN winds suppress $\Sigma_{\rm SFR}$ at all radii relative to the \nf~case, the higher fractional contribution of outflowing material to star formation can be indicative of local AGN feedback triggering of star formation, as star-forming gas is preferentially entrained within the outflowing medium.

\begin{figure}
\includegraphics[width = 1\columnwidth]{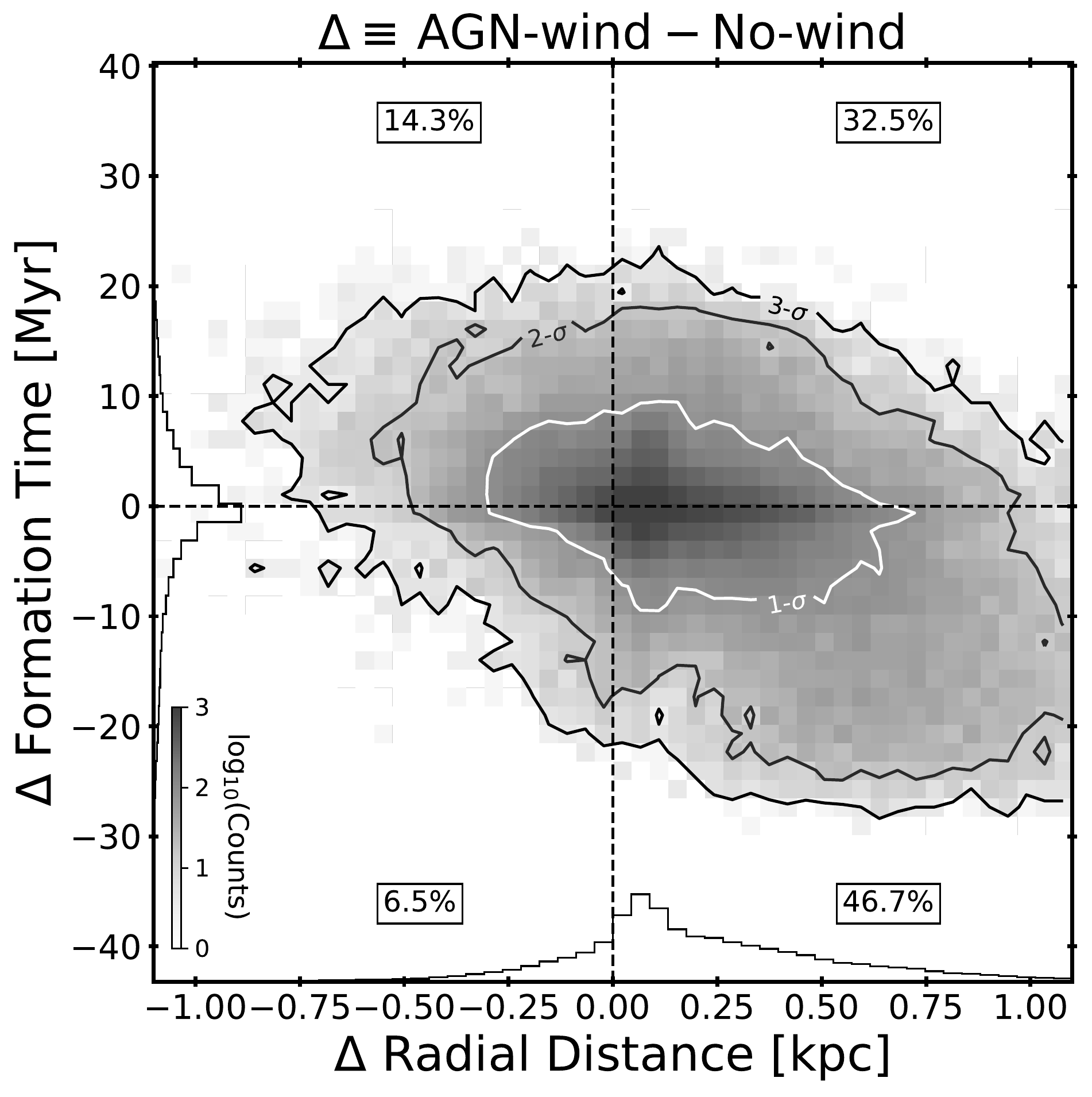}
\vspace*{-7mm}
\caption{\label{fig:deltaFTvsRD} Difference in formation time and radial distance for stars formed in the \mf~simulation relative to stars formed out of the same gas elements in the \nf~simulation. The two-dimensional distribution (and corresponding one-dimensional histograms) are shown for stars that formed in the \mf~simulation during : $0 \leq {\rm t}_{\rm form} < 30\,{\rm Myr}$. 
The radial distance to the centre is always measured at $\Delta t = 30$\,Myr. The contour lines highlight the regions that contains 1-$\sigma$, 2-$\sigma$, and 3-$\sigma$ of the distribution.
Dashed lines separate the distributions into four quadrants centred at [0,0], with the corresponding fraction of stellar mass indicated in each quadrant. Stars form preferentially farther out under the presence of AGN winds compared to the \nf~simulation (right quadrants), with a weak correlation between difference in formation time and distance.} 
\end{figure}

\subsection{Spatial and temporal shift in star formation}
\label{Spatial and temporal shift in star formation}

Since all simulations start from the same initial conditions ($\mathsection$\ref{subsec:initialconditions}), it is possible to track the evolution of the same Lagrangian mass elements (either gas or stars) across simulations using their unique particle identifiers.
Star formation is so efficient in the \nf~case that the majority of the gas particles that turned into stars in the \mf~case also formed stars in the \nf~simulation, which allows us to look further into the effect of AGN feedback on the stars that form in both cases. Figure~\ref{fig:deltaFTvsRD} quantifies the difference in the formation time and radial distance for stars formed in the \mf~simulation relative to the stars formed out of the same gas elements in the \nf~simulation. Here, we show the two-dimensional distribution for the differences in formation time and distance for stars that formed during the $30\,{\rm Myr}$ since the start of the AGN feedback phase (defined for the \mf~simulation).
This allows us to investigate whether stars that form in the presence of AGN winds do so at similar times/distances or not compared to the same stars in the absence of winds.  
For example, the top right quadrant corresponds to stars forming later in time and farther from the centre in the \mf~run, while the lower left quadrant represents stars forming earlier and closer than in the \nf~case.

The distribution in the difference of radial distances for stars that formed under the presence of AGN winds shows that $\sim$80\% of stars preferentially formed at a further distance compared to their \nf~counterpart. This may constitute indication of negative feedback, even for stars that manage to form in the presence of AGN feedback. When investigating the difference in formation times, we find some indication of earlier conversion of gas into stars for stars that form further out in the \mf~simulation (with AGN winds pushing ISM gas radially outward but locally triggering faster star formation) while preferentially delayed star formation for stars that form closer to the BH.
The complex interaction of AGN winds and ISM gas drives a weak anti-correlation between the difference in formation distance and the difference in formation time relative to the \nf~simulation.

\begin{figure}
\includegraphics[width = \columnwidth]{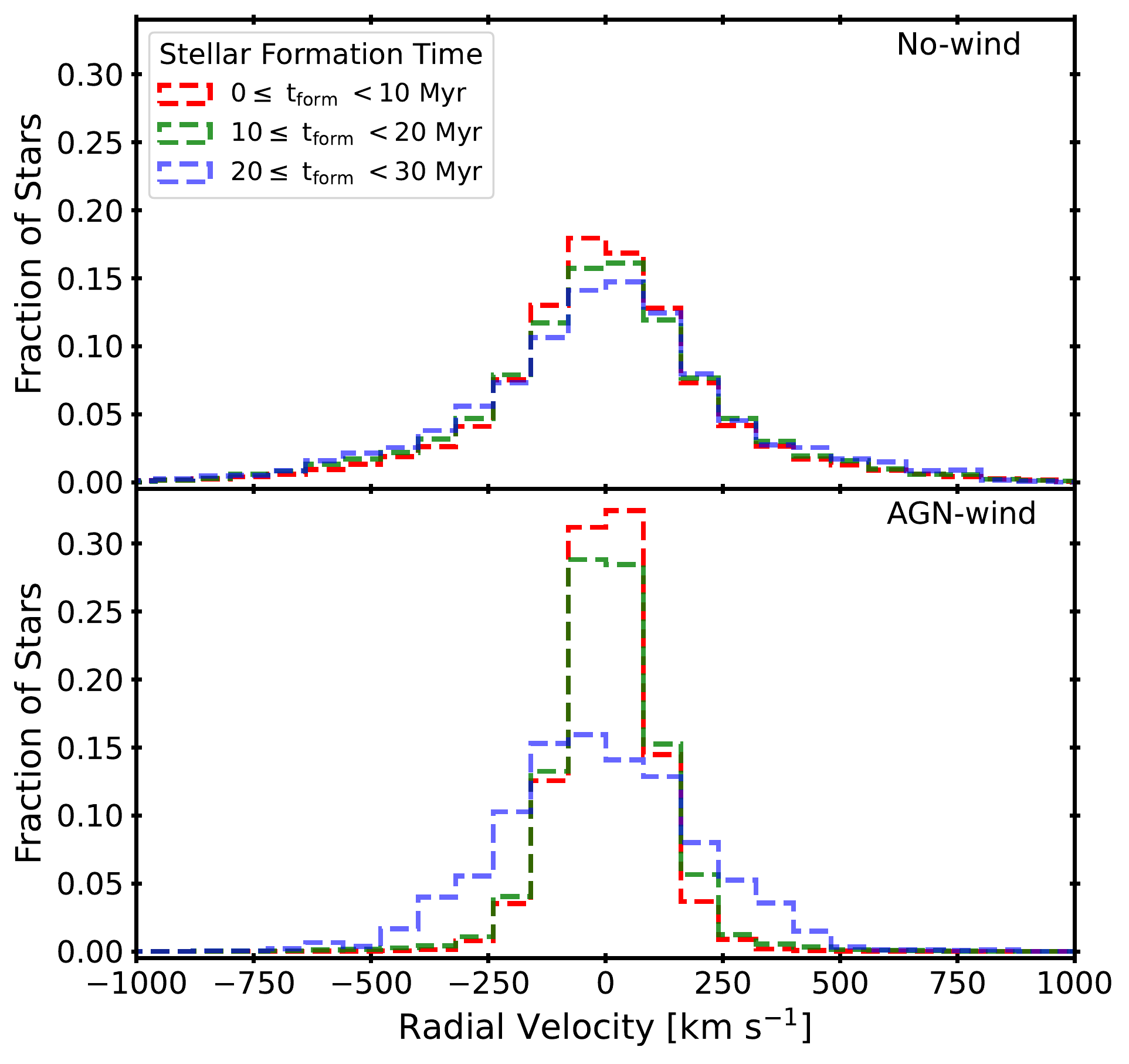}
\vspace*{-7mm}
\caption{\label{fig:RVhist} Radial velocity distribution for stars that formed in the \mf~simulation  within $\sim30$\,Myr (bottom panel), and the same star particles tracked in the \nf~simulation (top panel). Stars are separated according to their formation time in the \mf~simulation, as in Figure~\ref{fig:deltaFTvsRD}, with each colour corresponding to stars that formed in a given time interval as indicated. The radial velocity is always measured at $\Delta t=30$\,Myr.
}
\end{figure}

Figure~\ref{fig:RVhist} examines the radial velocities of the same sample of stars as in Figure~\ref{fig:deltaFTvsRD}, where we now show separate distributions for stars that formed in the \mf~simulation during three different time intervals: $0 \leq {\rm t}_{\rm form} < 10\,{\rm Myr}$, $10 \leq {\rm t}_{\rm form} < 20\,{\rm Myr}$, and $20 \leq {\rm t}_{\rm form} < 30\,{\rm Myr}$. In the \nf~simulation, the stellar radial velocity distributions are roughly symmetric and spread up to $\pm$1000\,$\kms$,~owing to the strong deepening of the nuclear gravitational potential in the absence of AGN feedback, with only minor differences depending on the stellar formation time. For comparison, the escape velocity at 1\,kpc increases from $\sim$620\,${\rm km}\,{\rm s}^{-1}$ at $\Delta t = 5\,{\rm Myr}$ to $\sim$700\,${\rm km}\,{\rm s}^{-1}$ at $\Delta t = 25\,{\rm Myr}$, while at 100\,pc the escape velocity increases from $\sim$750\,${\rm km}\,{\rm s}^{-1}$ to $\sim$1150\,${\rm km}\,{\rm s}^{-1}$ in the same time interval. In the absence of strong AGN or stellar-feedback driven winds that could potentially accelerate star-forming clumps, the large stellar velocities reached in the \nf~simulation can thus be easily explained by orbital dynamics in the ultra-dense nuclear stellar potential \citep{Angles-Alcazar2017c,Wellons2020,Parsotan2021}. 

In contrast, we find that new stars in the \mf~simulation are kinematically different in addition to forming further out compared to the \nf~case, exhibiting lower radial velocities than the same stars in the \nf~simulation. This is particularly the case for stars forming in the first $\sim 20\,$Myr, where $\sim$90\%~end up with radial velocities within $\pm$250\,$\kms$, owing to the lower gravitational potential. In this case, stars that form in the last $\sim$10\,Myr in the presence of winds show a broader velocity distribution compared to stars formed earlier, extending up to $\pm$500\,$\kms$ possibly due to the more coherent nature of star-forming structures dominated by either inflowing or outflowing material at higher velocities (Fig.~\ref{fig:RVvsRDall}). In any case, stellar radial velocities are a better reflection of the depth of the gravitational potential than of the velocity of AGN-driven winds. 

\section{Dependence on AGN wind efficiency} \label{Dependence on AGN wind efficiency}

\begin{figure}
\includegraphics[width = \columnwidth]{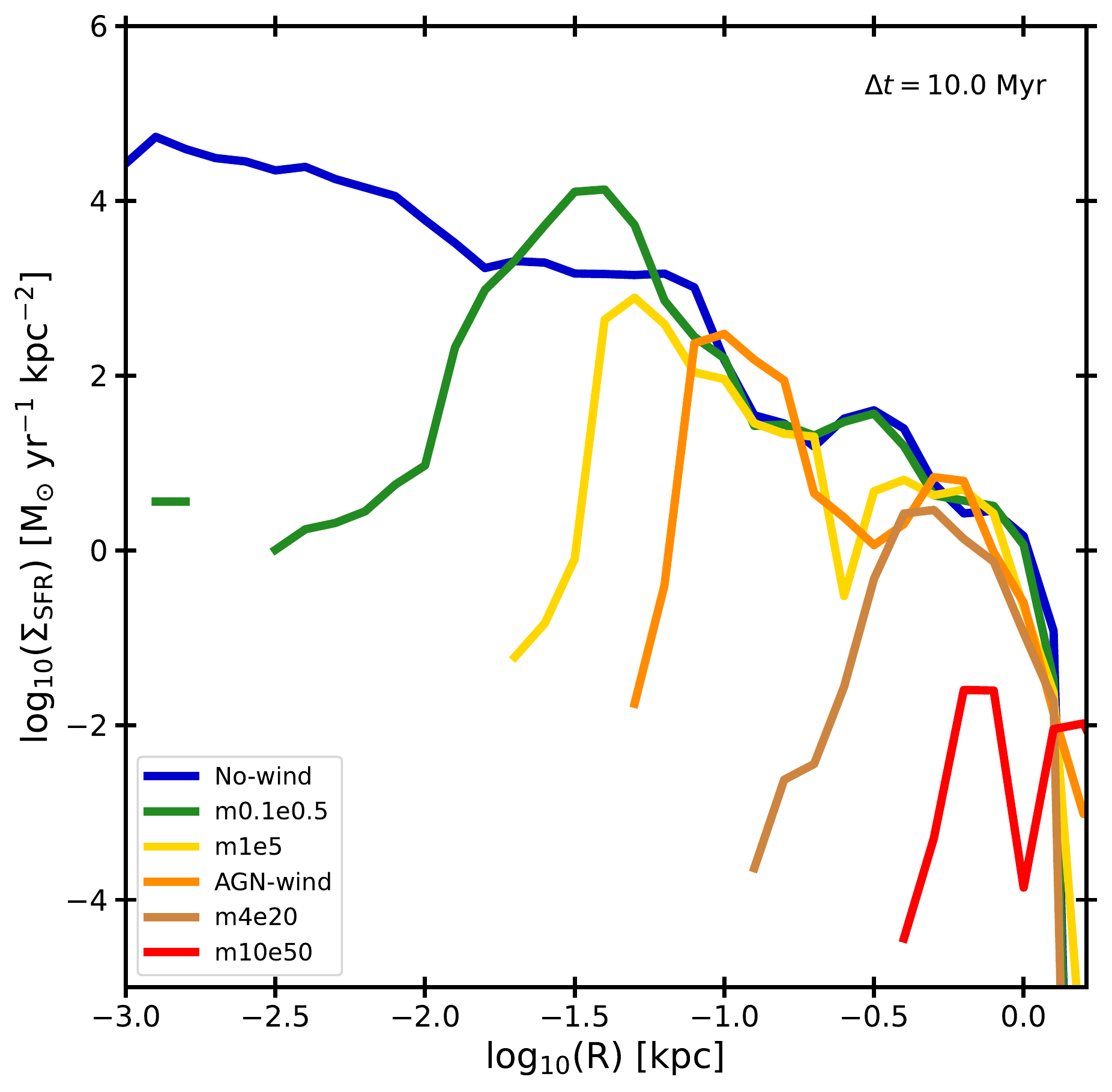}
\vspace*{-7mm}
\caption{\label{fig:sigmasfr_allsims} Radial profile of the total star formation rate surface density ($\Sigma_{\rm SFR}$) corresponding to $\Delta t=10$\,Myr since the start of the AGN phase, for various simulations with different AGN feedback strength.}
\end{figure}

Figure~\ref{fig:sigmasfr_allsims} shows the star formation rate surface density ($\Sigma_{\rm SFR}$) as a function of radial distance, similar to Figure~\ref{fig:mildfeedback_high} but at $\Delta t=10$\,Myr, for the \nf~and \mf~simulations, as well as four other simulations that vary in AGN feedback strength (see Table \ref{table:models}). At first glance, the hierarchy in feedback strength, from weakest ($\epsilon_{\rm k} = 0.5$\%; green) to strongest ($\epsilon_{\rm k} = 50$\%; red), can be clearly seen in the increasing suppression of the star formation rate surface density distribution relative to the \nf~case. The weakest AGN winds can barely open a central cavity of size $\sim25$\,pc during the first 10\,Myr, while the strongest winds very quickly evacuate the entire star-forming gas reservoir within the central kpc. Aside from seeing how $\Sigma_{\rm SFR}$ changes with different feedback efficiencies, we can also identify examples of local positive feedback. AGN winds in simulation \textbf{m0.1e0.5} can only suppress star formation within a small cavity, but the compression of gas in the edge of the cavity triggers star formation reaching a factor of ten higher $\Sigma_{\rm SFR}$ than in the \nf~simulation in the same radial range.

The top panel of Figure~\ref{fig:stellarmass_allsims} shows the stellar mass enclosed within $R<2$\,kpc as a function of time, excluding the mass found already in stars at the beginning of the AGN feedback phase at $t_{0}$, as in Figure~\ref{fig:stellarmass} but now comparing simulations with different AGN feedback strength.  
Similarly, the middle panel shows the stellar mass formed out of the initial star-forming gas reservoir at $t_{0}$ for each simulation, and the bottom panel shows the corresponding leftover star-forming gas.
The \textbf{m0.1e0.5} simulation shows very similar stellar mass growth as the \nf~case over time, indicating that AGN winds with kinetic feedback efficiency $\epsilon_{\rm k} = 0.5$\%~and energy injection rate $\dot{E}_{\rm w} \sim 6.29 \times 10^{44}\,{\rm erg\,{s}^{-1}}$ can only barely affect the star-forming gas reservoir of this massive galaxy. Increasing the AGN feedback strength by a factor of ten ($\epsilon_{\rm k} = 5$\%) has a clear negative effect in the overall stellar mass growth, suppressing the conversion of initially star-forming gas into stars and, more importantly, reducing the amount of new gas that can become star-forming, as seen in Figure~\ref{fig:stellarmass} for our fiducial \mf~simulation. In the strongest feedback cases (\textbf{m4e20} and \textbf{m10e50}), the total stellar mass enclosed within 2\,kpc increases initially by $\Delta M_{\star} \sim 10^{9}\,\Msun$ (during the first $\sim$5\,Myr) but then quickly decreases owing to the very strong suppression of subsequent star formation and the change in the gravitational potential due to the massive evacuation of gas from the galaxy driving the expansion of the stellar component.

\section{Discussion} \label{Discussion}

Our simulations suggest that powerful AGN winds have a global negative impact on the stellar mass growth of massive galaxies near their peak of star formation activity (Angl{\'e}s-Alc{\'a}zar et al., in prep.), in broad agreement with many previous cosmological simulations where AGN feedback prescriptions are calibrated to help regulate star formation in galaxies at the high mass end \citep{DiMatteo2005,Baldry2006,Bower2006,Dubois2012,Somerville&Dave2015,Choi2012,Choi2018,Dave2019,Wellons2023}. In the absence of AGN feedback, an ultra dense central starburst quickly develops at the time at which stellar feedback no longer becomes efficient enough to drive large-scale galactic winds \citep[][]{Angles-Alcazar2017c,Stern2021,Pandya2021,Byrne2023} and the simulated galaxy becomes overmassive and overcompact relative to observations \citep{Wellons2020,Parsotan2021}.
Our results indicate that sufficiently strong AGN winds (as would be produced by a MBH with $M_{\rm BH}= 10^{9}\,{\rm M}_{\odot}$ accreting at the Eddington rate with kinetic feedback efficiency $\epsilon_{\rm k} = 5$\%) can shut down star formation at this critical time, producing more realistic galaxy sizes and central stellar densities \citep[see][]{Cochrane2023}.
However, a factor of ten reduction in either $M_{\rm BH}$, $\epsilon_{\rm k}$, accretion rate relative to Eddington, or a combination of them (since these parameters are degenerate) can dramatically decrease the impact of AGN-driven winds, suggesting that other feedback channels, such as radiation pressure or cosmic rays (not included here) may be required to regulate massive galaxies \citep{Costa2018a,Costa2018b,Choi2018,Wellons2023}. 

AGN winds create a central cavity devoid of star-forming gas, as seen in previous studies \citep{Gabor2014,Curtis2016b,Hopkins2016,Richings2018,Costa2020,Torrey2020}, but can also penetrate into the galaxy disc to some extent, reducing the star formation rate surface density on all scales.
The coupling efficiency of AGN winds and the surrounding ISM decreases as the size of the cavity increases, with a larger fraction of input wind energy escaping along the polar direction \citep[][Angl{\'e}s-Alc{\'a}zar et al. in prep.]{Torrey2020}. The large, sustained gas inflow rate onto the galaxy yields a complex interaction between AGN winds in infalling structures, with dense star-forming clumps able to penetrate the cavity in some cases, and the geometric coupling efficiency of winds and ISM gas changing with time accordingly.  
Our results show that AGN winds reduce the amount of stars formed out of the initial star-forming gas reservoir but, more importantly, AGN winds are preventing new gas from becoming star-forming throughout the simulation.  

\begin{figure}
\includegraphics[width = \columnwidth ,valign=t]{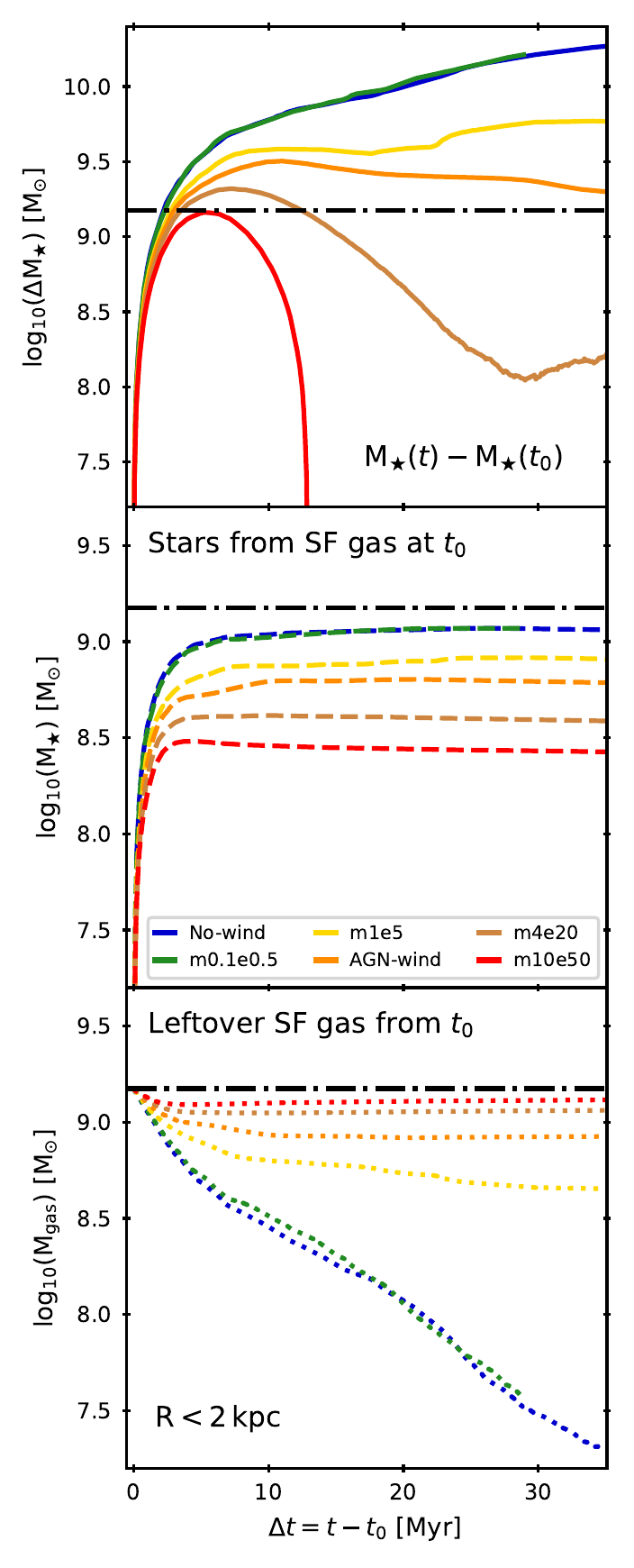}
\vspace*{-5mm}
\caption{\label{fig:stellarmass_allsims} Top: Stellar mass growth within the central 2\,kpc as a function of time for a variety of simulation runs with different AGN feedback strengths.  Solid lines represent the total mass of stars formed since $t_{0}$, i.e. excluding any pre-existing stars. Middle: The total mass in stars formed by initially star-forming gas at $t_{0}$ (dashed lines). Bottom: Total gas mass remaining from the original star-forming gas reservoir (dotted lines). The horizontal dash-dotted line (black) represents the initial total mass of star-forming gas available at the start of the simulations ($t_{0}$).}
\end{figure}

The impact of stellar and/or AGN feedback in galaxies has been generically considered as either {\it ejective}, removing gas directly from the ISM, or {\it preventive}, stopping gas from accreting into the ISM in the first place \citep{Somerville&Dave2015}.  Previous studies suggest that preventive AGN feedback is crucial in dwarf galaxies, where galactic winds reduce the accretion rate onto the galaxy \citep{Muratov2015,Hirschmann2016,Angles-Alcazar2017b, Hafen2019,Hafen2020,Mitchell2020,Tollet2019,Pandya2020}, and massive halos, where ``radio-mode'' or ``jet-mode'' AGN feedback has been proposed to prevent hot halo gas from cooling \citep{Bower2006,Gabor2015,Weinberger2017,Dave2019,Su2021}, while ejective feedback may be more prominent in gas-rich star-forming galaxies \citep{Somerville&Dave2015}.
Our analysis tracking the source of gas responsible for star formation in each simulation suggests that AGN winds are acting as both modes of negative feedback simultaneously, ejective and preventive \citep[see also][]{Grand2017,Irodotou2022}. 
Interestingly, preventive AGN-wind feedback (in the sense of preventing the replenishment of the star-forming gas reservoir) appears to be far more important than ejective feedback for suppressing the stellar mass growth of massive star-forming galaxies at their peak of activity (Figures~\ref{fig:stellarmass}~\&~\ref{fig:stellarmass_allsims}).

Our simulations show that AGN winds can also trigger star formation locally by compressing gas along the edge of the cavity, increasing the gas density and star formation efficiency locally compared to the same region in the simulation without AGN winds. This positive AGN feedback effect by gas compression is in qualitative agreement with previous analytic models and idealized simulations \citep[e.g.,][]{Silk2005,Gaibler2012,ishibashi2012,Zubovas2012,silk2013,zubovas2013,nayakshin2014,bieri2015,bieri2016,Dugan2017,zubovas2017}.
However, while many previous studies have argued that global positive AGN feedback plays a key role, and can even be the dominant star formation mode of gas-rich galaxies at high redshift \citep[e.g.,][]{Gaibler2012,silk2013,zubovas2013,bieri2015,bieri2016}, our simulations show that powerful AGN winds acting on a massive star-forming galaxy can only trigger a small amount of star formation compared to the overall negative effect.

Some studies propose that the dominant feedback effect, positive or negative, may depend on global host galaxy properties and/or the physical conditions in different parts of the galaxy. 
\citet{nayakshin2014} argues that AGN feedback has a negative effect in gas poor galaxies but a global positive effect in gas rich galaxies with AGN feedback accelerating the collapse of the cold ISM phase, in contrast with our results.
\citet{Gaibler2012} simulated the impact of AGN jets on idealized simulations of high-redshift galaxies, arguing that jet feedback suppresses star formation in the central region, forming a ring-like star-forming structure at the cavity boundary in qualitative agreement with our results, but they predict a strong increase in global SFR due to the external pressure enhancement throughout the galaxy disc produced by the jet cocoon \citep[see also][]{Sutherland2007,bieri2015,bieri2016,Dugan2017}. Recently, some simulations show that, enhancing the gas to higher densities near the jet axis may increase the likelihood of more star formation, although an overall increase in the turbulent velocity and kinetic energy results in suppressing star formation \citep{Mandal2021}.
Using idealized simulations of the impact of AGN outflows on the fragmentation rate of gas in turbulent spheres, \citet{zubovas2017} proposed that there is a critical AGN luminosity at which positive feedback dominates, with outflows being too efficient at removing gas at higher luminosities.   
Different galaxy properties, AGN feedback mechanisms, and efficiencies may thus play a role in the detailed balance of positive and negative feedback.  Our cosmological zoom-in simulations including a detailed treatment of star formation, stellar feedback, and hyper-refined AGN-driven winds propagating in a multi-phase ISM suggest that AGN feedback has either a minor global impact on massive star-forming galaxies (assuming low feedback efficiency) or net negative effects (for high feedback efficiency), but the contribution of positive feedback to star formation in galaxies under different conditions and/or redshifts should be investigated in future work. 

Signatures of local positive AGN feedback in our simulations include the spatial anti-correlation of wind-dominated regions and star-forming clumps, and higher local star formation efficiency in regions compressed by the winds, in qualitative agreement with observations \citep{cresci2015a,cresci2015b,Carniani2016,shin2019,Perna2020,Bessiere2022,Schutte2022}. Examples of high redshift galaxies where positive and negative AGN feedback may co-exist include SINFONI observations of quasars with extended outflows (traced by OIII) that appear to suppress star formation in a central cavity while triggering star formation (traced by H$\alpha$) along the edges of the outflow-dominated region \citep{cresci2015b,Carniani2016}, qualitatively similar to our simulated galaxy. However more recent observations with JWST suggest that H$\alpha$ may be powered by the central AGN rather than a signature of star formation \citep{cresci2023}. Examples in the low redshift universe include MUSE observations of Seyfert galaxies with similar spatial anti-correlations of outflow components and star-forming regions \citep{cresci2015a,shin2019}.  Combining MUSE and ALMA observations, \citet{shin2019} showed that regions of a star-forming ring structure interacting with a large-scale outflow in a nearby Seyfert 2 galaxy have higher SFR density but comparatively lower molecular gas content than other regions.  This may imply higher star formation efficiency by a factor $\sim$3--5 owing to positive AGN feedback, in qualitative agreement with our results.
Following a different approach, \citet{Bessiere2022} compared the spatial distribution of the young stellar populations ($<$100\,Myr) of the nearby Type II quasar Markarian 34 and the kinematics of the warm ionized outflows, finding a local enhancement of recent star formation coinciding with the outer edge of one side of the outflow, while increased turbulence and disrupted gas kinematics with no signs of recent star formation in the other side of the outflow. This provides further indication that positive and negative AGN feedback can coexist in galaxies, as suggested by our simulations. 
Nonetheless, simulated global galaxy SFRs are consistently higher in the absence of AGN winds, supporting scenarios where AGN feedback is globally negative (if anything) while star formation triggering represents a subdominant component.

In the absence of AGN winds, the star-forming gas reservoir contains roughly similar fractions of inflowing and outflowing components, reflecting the turbulent ISM dynamics prevalent throughout the simulation.  However, efficient AGN winds can significantly change the ISM gas kinematics. Simulations with AGN winds show stronger variability in the contributions of different dynamical components to the global SFR at later stages, varying from up to $\sim$90\%~of inflowing material to $\sim$90\%~of outflowing material over time.
Some observations suggest that outflows driven by AGN feedback contain star-forming gas within the outflow itself, which may constitute a strong manifestation of positive AGN feedback \citep{Santoro2016,maiolino2017,cresci2018,gallagher2019,RodriguezdelPino2019}.
We have identified some examples of outflowing star formation material, including a clear outflow component contributing $\gtrsim$70\%~of the total SFR (16\,$\Msunyr$ out of $\sim 20\,\Msunyr$) with velocities reaching up to $\sim$1000\,$\kms$. However, we do not see a consistent trend of outflowing gas dominating the SFR; the inflow component actually becomes more prevalent at later times, suggesting that gas pushed by the AGN winds has lower probability of forming stars except for short transitory phases.

The small overall amount of star formation in fast outflows suggests that positive AGN feedback does not contribute much to high velocity stars populating the stellar halo, possibly less than similar processes operating in galactic winds driven by stellar feedback \citep{Yu2020}.   
This is in contrast with some analytic models where gas swept outward by radiation pressure from the AGN efficiently forms outflowing stars that drive the size and structural evolution of massive galaxies \citep{ishibashi2012,Ishibashi2014}. Instead, the dominant structural effect of efficient AGN winds in our simulations is the suppression of the central starburst, with the consequent reduction in the nuclear stellar density and increase in the stellar effective radius of the galaxy \citep[][Angl{\'e}s-Alc{\'a}zar et al. in prep.]{Cochrane2023}.      
In a companion paper (Mercedes-Feliz et al., in prep.), we explore in detail the role of powerful AGN winds driving the formation of very dense stellar clumps in rare but extreme positive feedback events. 

Future work should consider the role of AGN winds on a larger sample of galaxies across different halo masses and redshifts, to investigate if the results presented here can be generalized beyond the specific galaxy conditions simulated here, and to evaluate whether host galaxy properties play a role in determining the efficiency of positive AGN feedback.
In addition, future simulations should consider the impact of AGN winds coupled to BH accretion self-consistently, unifying our wind particle spawning technique with 
the hyper-refinement scheme presented in \citet[][]{Angles-Alcazar2021} to increase the mass resolution dynamically as gas approaches the BH. 
The BH accretion rate is expected to decrease owing to AGN feedback self-regulation \citep[e.g.][]{DiMatteo2005,Choi2012,Hopkins2016,Habouzit2021}. Our results, assuming constant accretion rate, should thus represent an upper limit to the impact of accretion-driven winds for a given kinetic efficiency.

\section{Summary and conclusions} \label{Summary and Conclusions}

We have presented a detailed analysis of a set of high-resolution cosmological zoom-in simulations of a massive galaxy near the peak of star formation activity ($M_{\rm halo} \sim 10^{12.5}\,{\rm M}_{\odot}$ at $z\sim2$) to investigate the plausible positive versus negative effects of AGN feedback during a luminous quasar phase. Crucially, our simulations include resolved multi-phase ISM physics from the FIRE project \citep{Hopkins2018} and a novel implementation of hyper-refined AGN-driven winds that simultaneously captures their propagation and impact from the inner $\lesssim$10\,pc to CGM scales (Angl{\'e}s-Alc{\'a}zar et al., in prep.). Comparing simulations without MBH feedback and with different AGN feedback strength, our results can be summarized as follows:
\begin{itemize}

\item Strong AGN winds with 5\%~kinetic efficiency, powered by a MBH with mass $M_{\rm BH}= 10^{9}\,{\rm M}_{\odot}$ accreting at the Eddington rate, drive the formation of a central gas cavity of size $\sim$200\,pc and can dramatically reduce the star formation rate surface density across the galaxy disc in $\sim$30\,Myr, indicating that powerful quasar winds have a global negative impact on star formation.\\

\item Most of the star formation suppression relative to the control simulation without AGN winds is driven by a strong reduction in the amount of new gas that can become star-forming during a period of intense gas accretion onto the galaxy.  Direct ejection of the pre-existing star-forming gas reservoir at the beginning of the quasar phase only accounts for a small fraction of the overall reduction in stellar growth, suggesting that preventive feedback dominates over ejective feedback for strong AGN winds.\\

\item Reducing the kinetic efficiency by a factor of ten ($\epsilon_{\rm k} = 0.5$\%) shows no clear sign of global positive or negative feedback, with the star formation rate being barely affected,
while increasing the kinetic efficiency by a factor of ten ($\epsilon_{\rm k} = 50$\%) results in a complete blow out of the star-forming gas reservoir in $\lesssim$10\,Myr. \\

\item Compression of gas near the edge of the cavity by AGN winds can increase the local star formation rate surface density compared to the same radial range in the simulated galaxy without AGN winds. This local triggering of star formation is driven by the increased amount of gas accumulated at the edge of the cavity as well as the higher star formation efficiency of compressed gas. \\

\item The spatial anti-correlation of wind-dominated regions and star-forming clumps along with higher local star formation efficiencies constitute plausible signatures of local positive AGN feedback, in qualitative agreement with observations. However, the highest star formation efficiency reached across all simulations occurs in the absence of AGN winds, owing to the much larger central stellar mass surface density and the inability of stellar feedback to regulate star formation. \\

\item Besides the overall suppression of star formation, efficient AGN feedback drives a spatial and temporal shift in star formation.  Stars that do form under the presence of AGN winds tend to do so at larger radial distances compared to stars formed out of the same gas clumps in the absence of AGN winds.  Star formation also tends to occur earlier in time at the beginning of the quasar phase, with AGN winds triggering faster conversion of gas into stars in local regions, but comparatively later in time for stars forming after $\sim$20\,Myr of global negative impact of AGN winds. \\

\item The main impact of AGN feedback on the kinematics of new stars is the reduction of their typical radial velocity (either positive or negative). The strong deepening of the nuclear gravitational potential in the absence of AGN winds results in stellar radial velocities reaching up to $1000\,\kms$, while the suppression of the nuclear stellar density by AGN winds yields stellar radial velocities $\lesssim 500\,\kms$. \\

\item AGN winds tend to produce decoupling of dynamical components, with star-forming gas dominated by either inflowing or outflowing material in different regions. In some cases, the local contribution of outflowing material to star formation can exceed that of the inflow component by factors $>$10. However, the inflow component tends to dominate at later times, suggesting that gas radially accelerated by the winds has lower probability of forming stars except for short transitory phases.

\end{itemize}

In conclusion, our results suggest that positive and negative AGN feedback coexist in galaxies, where strong quasar winds can suppress global stellar mass growth while locally triggering a small amount of star formation. 
The simulations presented here do not support scenarios where positive AGN feedback is the dominant mechanism powering rapid star formation in galaxies.

\section*{Acknowledgements}
We thank the anonymous referee for constructive comments that helped improve the paper. The simulations were run on Flatiron Institute’s research computing facilities (Gordon-Simons, Popeye, and Iron compute clusters), supported by the Simons Foundation. We thank the Scientific Computing Core group at the Flatiron Institute for outstanding support.
Additional numerical calculations were run on the Caltech compute cluster “Wheeler,” allocations FTA-Hopkins supported by the NSF and TACC, and NASA HEC SMD-16-7592, and XSEDE allocation TG-AST160048 supported by NSF grant ACI-1053575.
JMF was supported in part by a NASA CT Space Grant Graduate Fellowship.
DAA acknowledges support by NSF grants AST-2009687 and AST-2108944, CXO grant TM2-23006X, Simons Foundation Award CCA-1018464, and Cottrell Scholar Award CS-CSA-2023-028 by the Research Corporation for Science Advancement.
SW was supported by an NSF Astronomy and Astrophysics Postdoctoral Fellowship under award AST2001905. 
CAFG was supported by NSF through grants AST-1715216, AST-2108230, and CAREER award AST-1652522; by NASA through grants 17-ATP17-006 7 and 21-ATP21-0036; by STScI through grants HST-AR-16124.001-A and HST-GO-16730.016-A; by CXO through grant TM2-23005X; and by the Research Corporation for Science Advancement through a Cottrell Scholar Award. JM is funded by the Hirsch Foundation. KS acknowledges support from the Black Hole Initiative at Harvard University, which is funded by grants from the John Templeton Foundation and the Gordon and Betty Moore Foundation, and support from Simons Foundation.

\section*{Data Availability}
The data supporting the plots within this article are available on reasonable request to the corresponding author. FIRE-2 simulations are publicly available \citep{Wetzel2023} at \url{http://flathub.flatironinstitute.org/fire}. Additional FIRE simulation data, including initial conditions and derived data products, are available at \url{https://fire.northwestern.edu/data/}. A public version of the GIZMO code is available at \url{http://www.tapir.caltech.edu/~phopkins/Site/GIZMO.html}.



\bibliographystyle{mnras}
\bibliography{main} 




\appendix
\section{Star formation efficiency per free-fall time}
\begin{figure*}
\centerline{\includegraphics[scale=.425,valign=t]{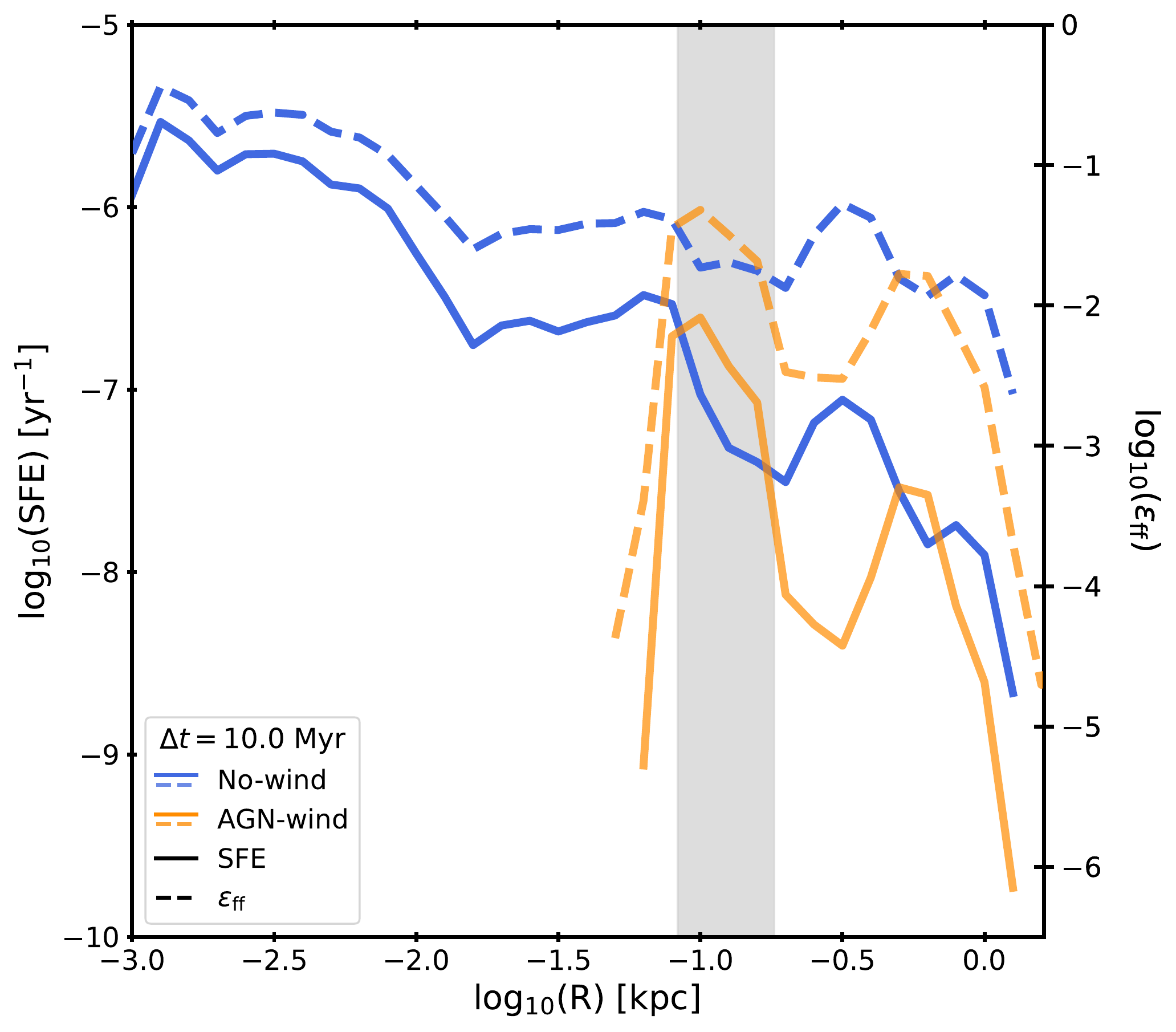} \includegraphics[scale=.425,valign=t]{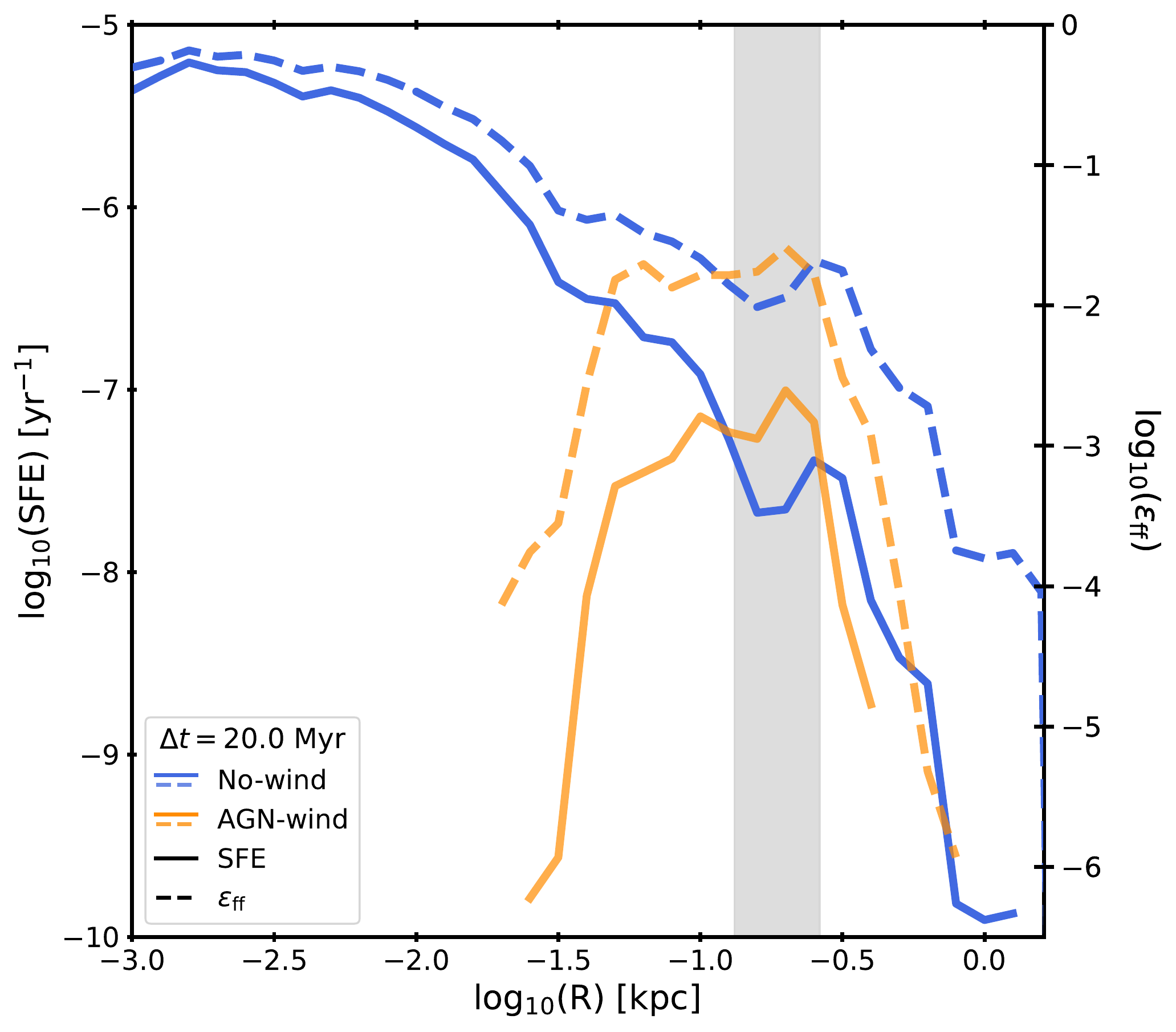}}
\vspace*{-2mm}
\caption{\label{fig:mildfeedback_sfehigh} Radial profile of the star formation efficiency (SFE; solid lines) and the SFE per free-fall time ($\epsilon_{\rm ff}$; dashed lines) at time $\Delta t = 10$\,Myr (left) and $\Delta t = 20$\,Myr (right) since the beginning of the quasar feedback phase for the \nf~(blue) and \mf~(orange) simulations.  The vertical grey band indicates the main region where SFE and $\epsilon_{\rm ff}$ are higher in the \mf~simulation compared to the \nf~case, which also corresponds to a region with higher \sigmasfr~in the presence of AGN winds.}
\end{figure*}

Figure~\ref{fig:mildfeedback_sfehigh} shows the radial profile of the star formation efficiency (SFE), as in Figure~\ref{fig:mildfeedback_sfehigh_mgm}, and the SFE per free-fall time ($\epsilon_{\rm ff}$) for two snapshots, $\Delta t= 10$\,Myr and $\Delta t= 20$\,Myr since the start of the quasar feedback phase. We compute \eff~$\equiv ({\rm SFR}/M_{\rm gas}) \times t_{\rm ff} = {\rm SFE}\times t_{\rm ff}$ for each gas element according to its mass $M_{\rm gas}$ and SFR, where the free-fall time is defined as $t_{\rm ff}=\sqrt{3\pi/32G\rho}$. We then compute the mass-weighted average over all gas in cylindrical radial bins.
The grey bands correspond to regions where \sigmasfr~is higher in the \mf~simulation compared to the \nf~case, which coincides with higher SFE and higher molecular gas mass surface density in the presence of AGN winds (Figure~\ref{fig:mildfeedback_sfehigh_mgm}). Here, we see that these regions also have higher effective star formation efficiency per free-fall time $\epsilon_{\rm ff}$, recovering similar results.
In the FIRE simulations, $\epsilon_{\rm ff}$ is a predicted quantity \citep{Hopkins2018}, with typical kpc scale-averaged efficiencies in the range \eff~$\sim 0.01-0.1$ for Milky Way-mass galaxies at $z=0$ \citep{Orr2018}, determined by stellar feedback self-regulation \citep{Ostriker2011,Hopkins2012,Faucher-Giguere2013}. Our simulations of a massive, high-redshift galaxy represent very different conditions either with or without AGN winds. In the \nf~simulation, we find that \eff~is significantly higher in the nuclear region because stellar feedback is no longer able to regulate star formation with such high stellar surface density; as expected, the average \eff~on kpc scales is much more similar to low-$z$ Milky Way-mass galaxies. In the \mf~case, the ultra-compact nuclear region does not develop, and the efficiency remains in the range \eff~$\sim 0.01-0.1$.


\bsp	
\label{lastpage}
\end{document}